\documentclass[12pt]{article}
\pdfoutput=1

\usepackage{draft,hyperref,tikz,cancel,subfig,float}
\usepackage[nosort]{cite}

\usetikzlibrary{snakes}
\usetikzlibrary{shapes.misc}
\usepackage{todonotes}
\usepackage{framed}
\usepackage{slashed}

\allowdisplaybreaks

\usetikzlibrary{shapes}
\tikzset{line/.style={line width=0.25mm},
curve/.style={line,smooth,tension=1},
->-/.style={decoration={
  markings,
  mark=at position #1 with {\arrow[>=stealth]{>}}},postaction={decorate}},
-<-/.style={decoration={
  markings,
  mark=at position #1 with {\arrow[>=stealth]{<}}},postaction={decorate}},
}

\begin{document}

%\reportnum{-3}{USTC-ICTS/PCFT-25-54}

\begin{titlepage}

\vspace*{-2.5cm} 
\noindent\hfill USTC-ICTS/PCFT-26-31\\\\

\begin{center}

\title{Defect Conformal Manifolds along RG Domain Walls between $\mathbb Z_N$-Parafermions and Minimal Models}
\author{Jin-Rui Zhang$^{\dagger}$, Jing-Hao Jin$^{\dagger}$, Ting-Kai Chen$^{\dagger}$, Jin Chen$^{\dagger,\,\natural,\,\flat}$}
\let\thefootnote\relax\footnotetext{$^\flat$ Corresponding author.}
\address{\small${}^\dagger$Department of Physics, Xiamen University, Xiamen, 361005, China}

\address{\small${}^\natural$Peng Huanwu Center for Fundamental Theory, Hefei, Anhui 230026, China}
%\vspace{-.1in}
%
\email{

jinruizhang@stu.xmu.edu.cn, 
jinghaojin520923@gmail.com, 
elychan@stu.xmu.edu.cn,
zenofox@gmail.com
}
\end{center}
\vfill

\begin{abstract}
We investigate the renormalization group (RG) domain walls interpolating between the $\mathbb{Z}_N$ parafermion theory (the critical $N$-state Potts model) and the Virasoro minimal model $\mathcal{M}_{N+1}$. These flows are genuinely non-perturbative and an explicit construction of Gaiotto type RG domain wall remains elusive. We bypass this limitation by employing a bottom-up approach centered on the emergence of ``phantom currents". By tracking the preserved non-invertible symmetries ($\mathfrak{so}(3)_N$) along the flow, we extract the exact spectrum of these currents localized on the defect. We demonstrate that the presence of a spin-1 phantom current allows the interface to be marginally deformed, dynamically generating a continuous defect conformal manifold. Furthermore, we show that an extra spin-2 operator, crucially as a $W^{(3)}$-algebra descendant of the spin-1 phantom current, rigidly constrains the UV-IR stress tensor mixing via the cluster decomposition principle. This algebraic framework enables the exact computation of the parameter-dependent transmission rate across the conformal manifold, which we observe strictly vanishes in the large-$N$ limit as a consequence of macroscopic target space collapse.
\end{abstract}
\vfill
\end{titlepage}

\tableofcontents

\newpage
\section{Introduction and Summary}
Conformal interfaces and boundaries are fundamental objects in two-dimensional conformal field theories (CFTs). By gluing two potentially distinct CFTs together while preserving a diagonal Virasoro algebra, conformal interfaces govern the transmission and reflection of energy and charges, providing deep insights into entanglement, dualities, and defect dynamics. Among them, Renormalization Group (RG) domain walls hold a particularly special status. Generated by a relevant deformation restricted to half of the spacetime, an RG domain wall encapsulates the exact operator mixing between the ultraviolet (UV) and infrared (IR) fixed points. Consequently, the highly non-trivial and intricate dynamics occurring in the strongly coupled regime along the RG flow are exactly encoded within these defects. However, such RG domain walls rarely admit a purely algebraic description. Generically, these RG domain walls are intrinsically irrational, even when the UV and IR endpoints are both rational CFTs (RCFTs). In the seminal paper \cite{Gaiotto:2012np}, a celebrated exact realization of such interfaces was given by Gaiotto, who constructed the RG domain walls between consecutive Virasoro minimal models\footnote{As we only consider unitary models throughout the paper, the notation $\mathcal M_{N+1}\equiv\mathcal M_{N+2,\,N+1}$ is used for brevity.} $\mathcal M_{N+2}^{UV}\rightarrow \mathcal M_{N+1}^{IR}$. The key insight in constructing this RG domain wall relies on the folding trick, embedding the product theory into an auxiliary theory $\mathcal T_{\mathcal B}$ with an extended chiral algebra $\mathcal B$,
\begin{align}
    \mathcal M_{N+2}^{UV}\otimes\overline{\mathcal M_{N+1}^{IR}}\longrightarrow \mathcal T_{\mathcal B} \,.
\end{align}
Within this extended theory, the RG domain wall can be elegantly realized as a rational Cardy brane equipped with an additional $\mathbb Z_2$ twist. This $\mathbb Z_2$ automorphism plays a vital role, as it intricately mixes the UV and IR stress tensors, thereby providing a completely algebraic description of the otherwise irrational RG interface. Related domain-wall constructions for RG flows between neighboring $W_3$ (equivalently, $A_2$) minimal models, including detailed analyses of UV--IR operator mixing, were developed in \cite{Poghosyan:2022mfw,Poghosyan:2023brb}.

The fundamental reason why the UV and IR theories can be embedded into a larger chiral algebra stems from the preservation of generalized (non-invertible) symmetries along the RG flow (for comprehensive reviews on generalized and non-invertible symmetries and their applications, see e.g., \cite{Fuchs:2002cm, Fuchs:2001qc,   Frohlich:2004ef, Fuchs:2012dt, Gaiotto:2014kfa, yujitasi, Bhardwaj:2017xup, Chang:2018iay, Choi:2021kmx, Komargodski:2020mxz, Thorngren:2019iar, Thorngren:2021yso, Kaidi:2022cpf,  Bhardwaj:2022yxj,Chang:2022hud, Kaidi:2023maf, Bhardwaj:2023wzd, Bhardwaj:2023ayw, Bhardwaj:2023kri, Chen:2023qnv, Shao:2023gho, Luo:2023ive, Cordova:2024iti, Chen:2025qub, Antinucci:2025fjp}). Specifically, the relevant deformation triggering the RG flow preserves a large class of non-invertible symmetries, realized as topological defect lines (TDLs) or Verlinde lines. These preserved topological lines are completely transparent to the RG domain wall, allowing them to freely pass through the interface. Crucially, these TDLs can terminate on specific defect operators. Upon applying the folding trick, the presence of these topological lines allows for a generalized orbifolding procedure, or say equivalently gauging the non-invertible symmetries. From a modern perspective, this gauging procedure corresponds to condensing these Verlinde lines into the vacuum \cite{Gaiotto:2020iye, Hu:2021qhm, Roumpedakis:2022aik, Choi:2022zal, Diatlyk:2023fwf, Chen:2024ulc}. This topological condensation enlarges the vacuum state, thereby rigorously defining the extended chiral algebra $\mathcal B$ of the auxiliary theory $\mathcal T_{\mathcal B}$. During this process, the defect operators residing at the endpoints of the condensed topological lines are absorbed into the new vacuum, becoming genuine local operators in the extended theory. These newly emerged local operators are precisely what have recently been termed phantom currents \cite{Antinucci:2025uvj, Furuta:2025ahl}.

The presence of such phantom currents on a conformal interface (or RG domain wall) is not merely a kinematic curiosity. Rather, these currents encode profound non-perturbative dynamical properties of the defect. As recently demonstrated by Copetti et al. \cite{Antinucci:2025uvj}, the existence of a spin-1 phantom current provides an exactly marginal operator localized on the domain wall. Deforming the interface by this operator preserves its conformal invariance, thereby generating a continuous moduli space of defects, aptly named a defect conformal manifold. A fascinating physical consequence of this marginal deformation is that macroscopic observables, such as the energy reflection and transmission rates across the domain wall, vary continuously as functions of the moduli parameter. On the other hand, the presence of a spin-2 phantom current, e.g. those emerging in Gaiotto's RG domain walls, plays a distinctly different but equally crucial role. In the unfolded picture, this spin-2 operator can participate in a highly non-trivial mixing between the stress tensor of the left (UV) theory $\mathcal T_{UV}$, and that of the right (IR) theory $\mathcal T_{IR}$, at the interface. As recently explored in \cite{Furuta:2025ahl}, this mixing implies that the transmission rate is rigidly constrained by the OPE data of the phantom current and stress tensors. In particular, it is observed that, in the unfolded picture, the two stress tensors $T_{UV}$ and $\overline T_{IR}$ in the UV and IR theories need to strictly satisfy the cluster decomposition principle, 
\begin{align}
    T_{UV}(z)\cdot{\overline T}_{IR}(w)\sim0\,.
    \label{eq:cluster_decomp}
\end{align}
This equation holds as an operator identity and therefore provides stringent constraints on the mixing coefficients among the spin-2 operators at the interface. Remarkably, the bottom-up approach allows one to determine macroscopic interface properties purely from local OPE data, completely bypassing the need for Gaiotto's non-trivial construction of the RG defect as a twisted Cardy state in the folded picture.

In this paper, we provide a concrete and non-trivial realization of this mechanism by investigating the RG flow from the $\mathbb Z_N$ parafermion $\mathcal P_N$ (or the critical $N$-state Potts model) to the Virasoro minimal model $\mathcal M_{N+1}$. Originally such a RG flow was discovered by Fateev and Zamolodchikov via the massless thermodynamic Bethe ansatz (TBA)\cite{Fateev:1985mm, Fateev:1987zh, Fateev:1990bf, Fateev:1991bv}. This flow is triggered by the least relevant operator, namely the parafermion operator and its charge conjugation, with conformal weight $h=1-\frac{1}{N}$,
\begin{align}
    \mathcal T_{\mathcal P_N}
   \ \xrightarrow[]{\text{least rel. def.}}
   \ \mathcal T_{\mathcal P_N}
    +
    \lambda_1
    \int d^2x\,
    \mathcal O_{\frac{N-1}{N}}(x)
    +
    \lambda_2
    \int d^2x\,
    \mathcal O^{\ast}_{\frac{N-1}{N}}(x)
    \quad\xrightarrow[]{\text{RG flow}}
    \quad
    \mathcal T_{\mathcal M_{N+1}}\,,
    \label{eq:def}
\end{align}
where ``$\ast$" is the $\mathbb Z_N$-charge conjugation operation. Unlike the well-known flows between consecutive minimal models, which become perturbative in the large-$N$ limit, this RG flow is genuinely non-perturbative, as the central charges of the UV and IR theories approach $c=2$ and $c=1$ respectively as $N\rightarrow\infty$. From the viewpoint of the Zamolodchikov metric, the present trajectory is correspondingly a long RG flow rather than a perturbatively short one. Related recent studies have uncovered broad families of symmetry-preserving, non-perturbative flows among Virasoro, superconformal, $\mathcal W$-algebra, and coset/parafermion models \cite{Nakayama:2024msv,Ambrosino:2025yug,Gaberdiel:2026sfg,Ambrosino:2026umb,Benedetti:2026drn}, to which the RG-wall methods developed here may also be applicable. Consequently, whether this RG domain wall admits a fully algebraic description, such as a Gaiotto-like twisted Cardy brane, remains a mystery. Nevertheless, the flow preserves a rich spectrum of phantom currents with integer spins
\begin{align}
   s(N) = \left\{ k^2 \;\middle|\; k \in \mathbb{Z}, \; 0 \le k \le \left\lfloor \frac{N}{2} \right\rfloor \right\}
\end{align}
allowing us to probe the interface dynamics purely from local OPE data. The presence of a spin-1 phantom current, $J=\phi_{UV}\,\bar\phi_{IR}$, suggests that the RG interface can be marginally deformed to generate a defect conformal manifold. Specifically, the 1d local defect operator induced by $J$ organizes the stress tensors and the spin-2 Virasoro primary 
\begin{align}
W^-\propto h_{IR}\partial \phi_{UV}\bar\phi_{IR}-h_{UV}\phi_{UV}\partial\bar\phi_{IR}
\end{align}
into a $\mathfrak{u}(1)$-multiplet, forcing $W^-$ to mix with the stress tensors $T_{UV}$ and $T_{IR}$ of the UV and IR theories at the interface. However, as shown in \cite{Furuta:2025ahl}, if $W^-$ is the sole spin-2 operator participating in the mixing, the cluster decomposition equation \eqref{eq:cluster_decomp} strictly demands $c_{UV}= c_{IR}$. This seemingly forbids the existence of a conformal manifold across our RG flow, where $c_{UV}\neq c_{IR}$ is inherently the case.

However, besides the operator $W^{-}$, there is an extra spin-2 operator due to the extended $W^{(3)}$-chiral algebra in the $\mathbb Z_N$ parafermion UV theory. More specifically, in the folded picture, the system harbors a spin-2 operator, 
\begin{align}
    X=W^{(3)}_{-1}\phi_{UV}\,\bar\phi_{IR}\,,
\end{align}
which is the $W$-algebra descendant of the spin-1 phantom current. Although there is no fundamental spin-2 phantom current in the preserved spectrum (the spins jump from 1 to 4), we can show that $X$ acts as a genuine, factorized Virasoro primary. It actively participates in the non-trivial mixing with the stress tensors at the interface alongside $W^-$. This leads to a compelling two-step physical picture: Firstly, introducing the parafermion fields deformation on a half-plane generates a rigid RG interface $\mathfrak D_N$ between the $\mathbb Z_N$-parafermion $\mathcal P_N$ and $\mathcal M_{N+1}$ model. Secondly, integrating the 1d local defect operator induced by the spin-1 phantom current along $\mathfrak D_N$ marginally deforms the interface, yielding a defect conformal manifold $\mathfrak D_N(\theta)$. The inclusion of the $W$-algebra descendant $X$ ensures that the cluster decomposition equation \eqref{eq:cluster_decomp} admits a continuous one-parameter family of non-trivial solutions for the mixing coefficients. This provides a highly non-trivial, explicit realization of the mechanism proposed in \cite{Antinucci:2025uvj} in a strictly $c_{UV}\neq c_{IR}$ case, demonstrating how the chiral algebra can dynamically sustain conformal manifolds on RG domain walls.

The remainder of this paper is organized as follows. In Section~\ref{sec:sym}, we first briefly review the coset construction of the $\mathbb Z_N$-parafermion $\mathcal P_N$ and the RG flow to minimal model $\mathcal M_{N+1}$. We then prove that the preserved fusion category along the RG flow is $\mathfrak{so}(3)_N$, i.e. the integer spin representation of $\mathfrak{su}(2)_N$, from which the spectrum of the preserved phantom currents can be read off. In Section~\ref{sec:RGwall}, we construct the RG domain wall and its conformal manifold between $3$-state Potts and $\mathcal M_{4}$ model. The explicit construction of such defect is made possible by the fact that the $3$-state Potts can be obtained via a $\mathbb Z_2$-orbifolding of the minimal model $\mathcal M_{5}$. Therefore one is able to establish the domain wall by stacking a topological interface onto the original Gaiotto wall between $\mathcal M_{5}$ and $\mathcal M_{4}$. In the rest of the section, we proceed to the general case $N$. Although an explicit construction of the RG domain wall for generic $N$ remains elusive, we show that the cluster decomposition equation will always admit a continuous one-parameter family of solutions, and thereby indicate the existence of the defect moduli. Finally, we compute the exact parameter-dependent transmission rate $\mathcal T_N(\theta)$ across this RG domain wall from $\mathcal P_N$ to $\mathcal M_{N+1}$,
\begin{align}
    \mathcal T_N(\theta)=\frac{2c_{UV\text{-}IR}}{c_{UV}+c_{IR}}=\frac{8(N+1)(1+\cos\theta)}{3(N+2)^2}\,,
\end{align}
as our main result, where $\theta$ is the moduli parameter of the defect conformal manifold along the RG domain wall $\mathfrak D_{N}(\theta)$. The explicit $N=3$ disk one-point functions and the corresponding UV--IR mixing coefficients are derived in App.~\ref{app:UV-IR-mixing}.

\section{Non-invertible Symmetries and Phantom Currents}
\label{sec:sym}
%\todo[inline]{2.1(JRZ) list facts on Pn and Mn+1, including central charge, labels, conformal weight, Smatrix; refer to app.A\\

%2.2(JRZ) LY fusion derivation delete\\

%2.3(JC)  embedding Ta->Tb; write more on phantom current (a few sentence), Interface simplify a bit.
%}

\subsection{Preliminaries on $\mathbb Z_N$-parafermions and minimal models}
In this section we collect the notation and modular data for the two coset RCFTs used below, see also in \cite{FATEEV1987644, Bouwknegt:1992wg}. A self-contained discussion on coset constructions, field identification, and selection rules is also given in App.~\ref{app:notation}.

The $\mathbb Z_N$ parafermion theory is realized as the coset
\begin{align}
    \mathcal P_N
    =
    \frac{\mathfrak{su}(2)_N}{\mathfrak{u}(1)_N},
    \qquad
    c(\mathcal P_N)
    =
    \frac{2(N-1)}{N+2}.
    \label{eq:parafermion-coset}
\end{align}
We label the primary fields by pairs
\begin{align}
    [t,s],
    \qquad
    t=1,\ldots,N+1,
    \qquad
    s\in \mathbb Z_{2N}.
\end{align}
The allowed labels obey the selection rule
\begin{align}
    t+s\in 2\mathbb Z+1,
\end{align}
and are subject to the field identification
\begin{align}
    [t,s]\sim [N+2-t,s+N],
    \qquad s \ \mathrm{mod}\ 2N.
    \label{eq:field-identification}
\end{align}
Thus the primary fields of $\mathcal P_N$ are equivalence classes of such labels. The conformal weight is
\begin{align}
    h^{\mathcal P_N}_{[t,s]}
    =
    \frac{t^2-1}{4(N+2)}
    -
    \frac{s^2}{4N}
    \quad \mathrm{mod}\;1 .
\end{align}

The modular $S$-matrix of the parafermion theory is
\begin{align}
    S^{\mathcal P_N}_{[t,s],[t',s']}
    =
    \sqrt{\frac{4}{N(N+2)}}\,
    \sin\!\left(\frac{\pi tt'}{N+2}\right)
    \exp\!\left(-\frac{i\pi ss'}{N}\right).
    \label{eq:parafermion-S}
\end{align}

The unitary Virasoro minimal model $\mathcal M_{k+2}$ is realized as
\begin{align}
    \mathcal M_{k+2}
    =
    \frac{\mathfrak{su}(2)_k\times \mathfrak{su}(2)_1}
         {\mathfrak{su}(2)_{k+1}},
    \qquad
    c(\mathcal M_{k+2})
    =
    1-\frac{6}{(k+2)(k+3)}.
\end{align}

Its primary fields are labeled by Kac labels with $d$ labels the representation of $\mathfrak{su}(2)_1$
\begin{align}
    [r,d,s],
    \qquad
    1\le r\le k+1,
    \qquad
    d=1,2,
    \qquad
    1\le s\le k+2,
\end{align}
with the field identification
\begin{align}
    [r,d,s]\sim [k+2-r,3-d,k+3-s].
\end{align}

The conformal weight is
\begin{align}
    h^{\mathcal M_{k+2}}_{[r,d,s]}
    =
    \frac{\big((k+3)r-(k+2)s\big)^2-1}
         {4(k+2)(k+3)} .
\end{align}

In these conventions, the modular $S$-matrix is
\begin{align}
    S^{\mathcal M_{k+2}}_{[r,d,s],[r',d',s']}
    =
    (-1)^{1+s r'+r s'}
    \sqrt{\frac{8}{(k+2)(k+3)}}\,
    \sin\!\left(\frac{\pi rr'}{k+2}\right)
    \sin\!\left(\frac{\pi ss'}{k+3}\right).
\end{align}

\subsection{Non-invertible symmetries in $\mathcal P_N$ and $\mathcal M_{{N+1}}$}
Non-invertible symmetries in two-dimensional CFTs are naturally realized by TDLs\cite{Chang:2018iay, Thorngren:2021yso}. Since a TDL can be freely deformed on the CFT as long as it does not cross local operator insertions, it provides a sharp diagnostic of which generalized symmetries survive under a relevant deformation. Let $\mathcal L$ be a topological line and $|0\rangle$ the unique vacuum. Its expectation value is defined by
\begin{align}
    \langle \mathcal L\rangle
    \equiv
    \langle 0|\mathcal L|0\rangle .
\end{align}
For a bulk primary operator $\phi$, we say that $\phi$ is neutral with respect to $\mathcal L$, or equivalently that $\mathcal L$ is transparent to $\phi$, if
\begin{align}
    \widehat{\mathcal L}|\phi\rangle
    =
    \langle \mathcal L\rangle |\phi\rangle .
    \label{eq:tdl-neutrality}
\end{align}
This condition states that the eigenvalue of $\phi$ under the action of the defect coincides with the eigenvalue of the vacuum. Therefore, inserting $\phi$ into the action does not obstruct the topological deformation of $\mathcal L$. A TDL satisfying \eqref{eq:tdl-neutrality} with respect to the perturbing operator is preserved along the RG flow.

For diagonal RCFTs, the TDLs are the usual Verlinde lines. They are labelled by bulk primaries and act diagonally on the Hilbert space as
\begin{align}
    \mathcal L_i |\phi_j\rangle
    =
    \frac{S_{ij}}{S_{0j}}|\phi_j\rangle ,
    \label{eq:verlinde-line-action}
\end{align}
where $S_{ij}$ is the modular $S$-matrix and $0$ denotes the vacuum primary. Equivalently, in the three-dimensional topological field theory description, bulk primaries are represented by Wilson lines piercing the two-dimensional surface, while TDLs are Wilson lines lying on the surface. The eigenvalue in \eqref{eq:verlinde-line-action} is then the normalized Hopf-link invariant obtained when the defect line encircles the Wilson line associated with $\phi_j$. For non-diagonal theories, the one-to-one correspondence between bulk primaries and TDLs need not hold. Nevertheless, the fusion category of topological lines still encodes the generalized symmetry data of the CFT and therefore imposes non-trivial constraints on possible RG flows.

We now apply this criterion to $\mathcal P_N$. Its most relevant deformation is generated by a primary field of conformal weight
\begin{align}
    h=1-\frac{1}{N}.
\end{align}
It may be represented by the fields $\phi_{[1,2]}$ and $\phi_{[1,2N-2]}$, or by an appropriate real linear combination of them. A topological line $\mathcal L_{[t,s]}$ is preserved by this deformation precisely when its eigenvalue on the perturbing field agrees with its eigenvalue on the vacuum. Using \eqref{eq:verlinde-line-action} and \eqref{eq:parafermion-S}, this condition reduces to the simple phase constraint
\begin{align}
    \exp\!\left(-\frac{2\pi i s}{N}\right)=1 .
\end{align}
Hence
\begin{align}
    s=0
    \quad \text{or}\quad
    s=N
    \qquad
    \mathrm{mod}\ 2N .
    \label{eq:preserved-para}
\end{align}
Because the two choices are related by the field identification \eqref{eq:field-identification}, we may choose representatives with $s=0$. The parity condition then requires $t$ to be odd. Therefore the preserved topological lines are represented by
\begin{align}
    \mathcal L_{[t,0]},
    \qquad
    t\in 2\mathbb Z+1 .
    \label{eq:preserved-lines}
\end{align}
The other possibility is related by the field identification
\begin{align}
    \mathcal L_{[t_k,N]}\sim \mathcal L_{[N+2-t_k,0]}.
    \label{eq:TDL-id}
\end{align}
We next determine the fusion rules among these preserved lines. Let
\begin{align}
    \ell=t-1 .
\end{align}
Since the preserved representatives have $t\in 2\mathbb Z+1$, they correspond precisely to the even-$\ell$, or integer-spin, labels of $\mathfrak{su}(2)_N$. Therefore
\begin{align}
    \mathcal L_{[\ell_i+1,0]}\times\mathcal L_{[\ell_j+1,0]}
    =
    \bigoplus_{\substack{
        \ell=
        |\ell_i-\ell_j|\\
        \mathrm{step}\;2
    }}^{
        \min(\ell_i+\ell_j,\;2N-\ell_i-\ell_j)
    }
    \mathcal L_{[\ell+1,0]},
    \qquad
    \ell_i,\ell_j,\ell\in 2\mathbb Z .
    \label{eq:SO3N-fusion}
\end{align}
Equivalently,
\begin{align}
    N_{\ell_i,\ell_j}^{\;\;\ell_k}
    =
    \sum_{\substack{
        \ell=
        |\ell_i-\ell_j|\\
        \mathrm{step}\;2
    }}^{
        \min(\ell_i+\ell_j,\;2N-\ell_i-\ell_j)
    }
    \delta_{\ell_k,\ell},
    \qquad
    \ell_i,\ell_j,\ell_k\in 2\mathbb Z .
\end{align}
Here we summarize the result, while the detailed calculation is given in App.~\ref{app:sym}. These are precisely the fusion rules of the integer-spin subcategory of \(\mathfrak{su}(2)_N\), usually denoted by
\begin{align}
    \mathcal C_{\mathrm{pres}}
    \simeq
    \mathfrak{so}(3)_N .
\end{align}
Related appearances of the same category in parafermionic constructions were recently found in the parafermionization of the Monster CFT, where a $\mathrm{Rep}(\mathfrak{so}(3)_p)$ symmetry was argued to emerge for odd prime $p$ \cite{Honda:2026bjy}. In physical terms, the relevant perturbation does not preserve the full parafermionic defect category, but it does preserve the non-invertible subcategory generated by the lines \eqref{eq:preserved-lines}. This preserved category must be matched, or embedded, in the generalized symmetry category of the IR fixed point \cite{Benedetti:2026drn}.

Let us illustrate the preservation of the deformation symmetry $\mathcal P_3$. Following the notation used above, the relevant deformation is generated by the two fields of conformal weight $h=2/3$,
\begin{align}
    \mathcal T_{\mathcal P_3}
    \longrightarrow
    \mathcal T_{\mathcal P_3}
    +\lambda_1
    \int d^2x\,
    \mathcal O_{\frac23}(x)
    +
    \lambda_2
    \int d^2x\,
    \mathcal O^\ast_{\frac23}(x).
\end{align}
By the general selection rule, the topological lines preserved by this deformation are the lines labeled by $\mathcal L_{[t,0]}$ with odd $t$. For $N=3$ this leaves
\begin{align}
    \{\mathcal L_{[1,0]},\mathcal L_{[3,0]}\}\simeq \text{LY}.
\end{align}
Thus, in this example, the general deformation-preserved $\mathfrak{so}(3)_N$ symmetry reduces at $N=3$ to a preserved LY non-invertible symmetry.

This preserved LY symmetry strongly constrains the possible infrared behavior. A theory with an unbroken LY symmetry cannot end in a trivially gapped phase with a single featureless ground state. Hence the LY-preserving deformation has two possible types of infrared behavior: either it becomes gapped in a non-trivial way, or it remains gapless with a single conformal vacuum. The massless TBA analysis \cite{Fateev:1991bv} selects a special integrable ray in the coupling space of this perturbation. In the standard conjugate normalization of the two perturbing operators, the gapless branch is characterized by equal magnitudes of the two couplings and by the relative phase
\begin{align}
    |\lambda_1|=|\lambda_2|,\qquad
\frac{\lambda_1}{\lambda_2}=e^{2\pi i/N},
\end{align}
In the present case $N=3$. Therefore the massless condition becomes
\begin{align}
    \frac{\lambda_1}{\lambda_2}=e^{2\pi i/3},
\end{align}
Thus the RG trajectory considered here is not a generic two-parameter LY-preserving deformation, but the $N=3$ massless integrable ray. Since the preserved LY symmetry forbids a trivially gapped phase with a single featureless ground state, this identifies the deformation as a gapless LY-preserving flow rather than the massive parafermionic branch.

The possible conformal endpoints are then further restricted by $c$-theorem \cite{Zamolodchikov:1987ti}. Since
\begin{align}
    c(\mathcal P_3)=\frac45,
\end{align}
any non-trivial unitary minimal-model endpoint must have smaller central charge. Among the relevant candidates below $c=4/5$, this leaves $\mathcal M_3$ and $\mathcal M_4$. Defect category of the critical Ising model is of $\text{TY}_2$ type and does not contain the required LY line. Therefore it cannot match the non-invertible symmetry preserved along the $\mathcal P_3$ deformation. By contrast, the tricritical Ising model $\mathcal M_4$ contains an LY sector. Hence the preserved symmetry can be matched in the RG flow,
\begin{align}
    \mathcal P_3
    \longrightarrow
    \mathcal M_4\,.
\end{align}

\subsection{Embedding, decomposition, and phantom currents}
By the folding trick, an RG domain wall between the UV theory $\mathcal T_{UV}$ and the IR theory $\mathcal T_{IR}$ can be equivalently regarded as a conformal boundary condition in the folded product theory. In the case we discuss, it is
\begin{align}
    \mathcal T_\mathcal A=\frac{\mathfrak{su}(2)_N}{\mathfrak{u}(1)_N}\otimes \frac{\mathfrak{su}(2)_{N-1}\times\mathfrak{su}(2)_1}{\mathfrak{su}(2)_N}
    \ \longrightarrow\ 
    \mathcal T_{\mathcal B}=\frac{\mathfrak{su}(2)_{N-1}\times\mathfrak{su}(2)_1}{\mathfrak{u}(1)_N}\,.
\end{align}
The boundary condition becomes elementary only after extending the chiral algebra of $\mathcal T_{\mathcal A}$ to a larger rational algebra. We denote the corresponding extended theory by $\mathcal T_{\mathcal B}$.

Let $a,b,\ldots$ label irreducible sectors of the original folded theory $\mathcal T_{\mathcal A}$, and let $\mu,\nu,\ldots$ label irreducible sectors of the extended theory $\mathcal T_{\mathcal B}$. Upon restricting a representation of $\mathcal T_{\mathcal B}$ to the smaller chiral algebra of $\mathcal T_{\mathcal A}$, its character decomposes as
\begin{align}
    \chi^{\mathcal B}_{\mu}(\tau)
    =
    \sum_{a\in \mathcal A}
    n_{\mu}^{a}\,
    \chi^{\mathcal A}_{a}(\tau) \, ,
    \label{eq:branching_character}
\end{align}
where $n_{\mu}^{a}\in \mathbb Z_{\geq 0}$ is the branching multiplicity. Modular covariance requires the branching matrix to intertwine the modular $S$-matrices of the two theories:
\begin{align}
    \sum_{a}
    n_{\mu}^{a}\,
    S^{\mathcal A}_{ab}
    =
    \sum_{\nu}
    S^{\mathcal B}_{\mu\nu}\,
    n_{\nu}^{b} \, .
    \label{eq:branching_S_intertwining}
\end{align}
This relation will be used repeatedly below to convert objects in the extended boundary theory into those in the folded product theory.

For the RG flow from the $\mathbb Z_N$ parafermion theory $\mathcal P_N$ to the minimal model $\mathcal M_{N+1}$, the folded theory is of the form
\begin{align}
    \mathcal T_{\mathcal A}
    =
    \mathcal P_{N}\otimes\overline{\mathcal M_{N+1}} \, .
\end{align}
The relevant extension can be described by the branching of the extended primaries into the product theory. Schematically, and suppressing the standard selection rules and field identifications, we write
\begin{align}
    \phi^{\mathcal B}_{[t,d,s]}
    =
    \sum_{r}
    \left(
    {\phi}^{\mathcal P_N}_{[r,s]}
    \otimes
    \bar\phi^{\mathcal M_{N+1}}_{[t,d,r]}    
    \right)\,.
    \label{eq:PN_branching}
\end{align}
The index $r$ runs over the admissible $\mathfrak{su}(2)_N$ labels. Equation \eqref{eq:PN_branching} is the concrete realization of \eqref{eq:branching_character} in the present class of flows.

We now recall how the corresponding interface is written in the folded Ishibashi basis. For every pair $(\mu,a)$ with nonzero multiplicity $n_{\mu}^{a}$, we introduce Ishibashi states
\begin{align}
    |\!| \mathcal A;\mu,a;i \rangle\!\rangle ,
    \qquad
    i=1,\ldots,n_{\mu}^{a} \, ,
\end{align}
where the extra label $i$ resolves possible multiplicities in the branching. A Cardy boundary condition $x$ of the extended theory $\mathcal T_{\mathcal B}$ defines an interface in the unfolded picture:
\begin{align}
    \mathcal I_x
    =
    \sum_{\mu,a}
    \sum_{i=1}^{n_{\mu}^{a}}
    \frac{S^{\mu,a}_{x}}{S^{\mathcal A}_{1a}}
    \,
    |\!| \mathcal A;\mu,a;i |\!| .
\end{align}
A particularly important interface is
\begin{align}
    \mathcal I_1
    =
    \sum_{\mu,a}
    \sum_{i=1}^{n_{\mu}^{a}}
    \sqrt{
    \frac{S^{\mathcal B}_{1\mu}}
         {S^{\mathcal A}_{1a}}
    }
    \,
    |\!| \mathcal A;\mu,a;i |\!| .
    \label{eq:identity_interface}
\end{align}
The interface $\mathcal I_1$ plays two complementary roles, which together motivate the bottom-up strategy of this paper. First, $\mathcal I_1$ specifies which operators of the folded theory $\mathcal T_\mathcal A$ condense into the new vacuum module of the extended theory $\mathcal T_\mathcal B$. The Ishibashi sum in eq.~\eqref{eq:identity_interface} is supported on pairs $(\mu,a)$ with non-vanishing branching multiplicity $n^a_\mu$, and restricting to the vacuum sector $\mu=1$ leaves precisely the multiplicities $\{n_1^a\}$. These multiplicities catalogue the irreducible $\mathcal A$-sectors that contribute to the extended vacuum of $\mathcal B$; Secondly, fusing $\mathcal I_1$ with its orientation-reversed counterpart $\widetilde {\mathcal I}_1$ collapses the Ishibashi sum into a sum of topological defect lines of $\mathcal T_\mathcal A$, 
\begin{align}
    \widetilde{\mathcal I}_1\circ\mathcal I_1
    =
    \sum_{a}
    n_{1}^{a}\,
    \mathcal L_{a} \, .
    \label{eq:interface_fusion_TDL}
\end{align}
Here $\mathcal L_a$ denotes the topological defect line of $\mathcal T_{\mathcal A}$ labeled by the sector $b$. The last equality follows directly from the modular intertwining relation \eqref{eq:branching_S_intertwining}. More importantly, the collection 
\begin{align}
\mathcal C_{\rm pres}=\left\{\mathcal L_a\,|\,n_1^a>0\right\}    
\end{align}
is exactly those TDLs penetrating the RG domain wall and coexist in both $\mathcal T_{UV}$ and $\mathcal T_{IR}$ in the unfolded picture. They form a closed fusion subcategory of the symmetry category of $\mathcal T_{\mathcal A}$, equipped with the structure of a commutative Frobenius algebra. The chiral extension $\mathcal A\rightarrow \mathcal B$ is realized as the topological condensation (equivalently, the generalized gauging) of this subcategory inside $\mathcal T_\mathcal A$:
\begin{align}
    \mathcal T_\mathcal B=\mathcal T_\mathcal A/\mathcal C_{\rm pres}\,.
\end{align}
Condensing the lines $\mathcal L_a$ enlarges the vacuum, and the defect operators previously living at the endpoints of these lines are absorbed into the new vacuum as genuine local fields of $\mathcal T_\mathcal B$. These newly local fields are exactly the phantom currents. From eq.~\eqref{eq:PN_branching}, one finds that
\begin{align}
    \phi^{\mathcal B}_{[1,1,0]}
    =
    \sum_{k=0}^{\lfloor N/2\rfloor}
    \left(
    {\phi}^{\mathcal P_N}_{[2k+1,0]}
    \otimes
    \bar\phi^{\mathcal M_{N+1}}_{[1,1,2k+1]}    
    \right)\,.
    \label{eq:Bvac}
\end{align}
where the conformal weights of ${\phi}^{\mathcal P_N}_{[2k+1,0]}$ and $\phi^{\mathcal M_{N+1}}_{[1,1,2k+1]}$ are
\begin{align}
    h_{[2k+1,0]}^{\mathcal P_N}=\frac{k(k+1)}{N+2}\,,\qquad
    h^{\mathcal M_{N+1}}_{[1,1,2k+1]}=k^2-\frac{k(k+1)}{N+2}\,,
\end{align}
indicating that the vacuum module $\phi^{\mathcal B}_{[1,1,0]}$ of $\mathcal T_\mathcal B$ contains a rich spectrum of phantom currents with integer spins
\begin{align}
   s(N) = \left\{ k^2 \;\middle|\; k \in \mathbb{Z}, \; 0 \le k \le \left\lfloor \frac{N}{2} \right\rfloor \right\}\,.
\end{align}
The two viewpoints are dual descriptions of the same data: the operators entering the new vacuum of $\mathcal T_\mathcal B$ (first viewpoint) are precisely the local fields created by line condensation in $\mathcal T_\mathcal A$ (second viewpoint), and the controlling multiplicities $\{n_1^a\}$ are common to both. In what follows we use whichever description is more convenient, the operator/branching description for spectral statements, and the line/condensation description for symmetry-theoretic ones.

The interface $\mathcal I_1$ introduced above is, however, only kinematic data: it specifies the operator content compatible with the chiral extension $\mathcal A\rightarrow\mathcal B$, but does not by itself realize the RG domain wall. A genuine RG wall, when it admits a fully algebraic description as a rational Cardy brane of $\mathcal T_\mathcal B$, requires the additional input of an automorphism $\sigma$ of the chiral algebra $\mathcal B$. The $\sigma$-twist mixes the UV and IR stress tensors non-trivially at the wall and supplies the dynamical content that $\mathcal I_1$ alone cannot encode. Concretely, $\sigma$ defines a $\sigma$-twisted Cardy state in $\mathcal T_\mathcal B$; fusing this twisted brane with the interface $\mathcal I_1$ yields the Gaiotto RG brane in the folded picture $\mathcal T_\mathcal A$. For the consecutive minimal model flows $\mathcal M_{N+2} \rightarrow \mathcal M_{N+1}$, the relevant $\sigma$ is the $\mathbb Z_2$ automorphism that swaps the two $\mathfrak{su}(2)_1$ numerator factors of Gaiotto's auxiliary theory $\mathcal T_\mathcal B^{\,\prime}$, and the resulting wall reproduces the expected UV--IR operator mixing.

For the parafermion flow $\mathcal P_N \rightarrow \mathcal M_{N+1}$ studied here, the corresponding automorphism is not known yet for general $N$: the natural auxiliary theory $\mathcal T_\mathcal B=\mathfrak{su}(2)_{N-1} \times \mathfrak{su}(2)_1/\mathfrak{u}(1)_N$ does not carry the manifest swap symmetry that powers Gaiotto's construction. The lone accessible case is $N = 3$, where the well-known isomorphism $\mathcal P_3 \simeq \mathcal M_5 / \mathbb Z_2$ reduces the problem to a topological manipulation: fusing Gaiotto's $\mathcal M_5 \rightarrow \mathcal M_4$ wall with the $\mathbb Z_2$-orbifold topological interface relating $\mathcal M_5$ and $\mathcal P_3$ produces the desired RG domain wall. We carry out this construction explicitly in the first part of next section. For general $N$ no such reduction is available, and a fully algebraic realization of the RG wall remains an open problem. Therefore, in the second part of next section, we adopt a complementary, bottom-up strategy: we exploit the local OPE data of the phantom currents to extract its macroscopic observables directly. As we shall show, the cluster decomposition condition \eqref{eq:cluster_decomp}, together with the spin-1 and spin-2 phantom currents identified above, is sufficient to determine the transmission rate of the wall and the dimension of its defect conformal manifold — without committing to an explicit construction of the wall itself.

\section{RG Domain Wall between $\mathbb Z_N$-parafermion and $\mathcal M_{N+1}$}
\label{sec:RGwall}

\subsection{An appetizer: $\mathbb Z_3$-parafermion to $\mathcal M_{4}$}
The $\mathbb Z_3$-parafermion model $\mathcal P_3$ is well understood to be obtained from a $\mathbb Z_2$-orbifolding of the minimal model $\mathcal M_{5}$. To construct the RG domain wall between $\mathcal P_3$ and $\mathcal M_{4}$, let us first review the construction of the wall between $\mathcal M_{5}$ and $\mathcal M_{4}$. In \cite{Gaiotto:2012np}, Gaiotto realized that the folded theory $\mathcal M_{5}\otimes \overline{\mathcal M_{4}}$ can be embedded in the auxiliary theory $\mathcal T^{\,\prime}_{\mathcal B}$,
\begin{align}
    \mathcal M_{5}\otimes \overline{\mathcal M_{4}}=\frac{\mathfrak{su}(2)_3\times\mathfrak{su}(2)_1}{\mathfrak{su}(2)_4}\times \frac{\mathfrak{su}(2)_2\times\mathfrak{su}(2)_1}{\mathfrak{su}(2)_3}
    \ \longrightarrow\ 
    \mathcal T^{\,\prime}_{\mathcal B}=\frac{\mathfrak{su}(2)_2\times\mathfrak{su}(2)_1\times\mathfrak{su}(2)_1}{\mathfrak{su}(2)_4}\,,
\end{align}
In $\mathcal T^{\,\prime}_{\mathcal B}$, the vacuum character is
\begin{align}
    \chi_{\rm vac}^{\mathcal B'}=q^{-\frac{c_{UV}+c_{IR}}{24}}\left(1+3q^2+\cdots\right)\,.
    \label{eq:vacuum-character-Bprime}
\end{align}
Here $q=e^{2\pi i\tau}$ keeps track of the holomorphic conformal weight above the vacuum. Since $L_{-1}|0\rangle=0$, the coefficient of $q^2$ counts independent spin-two states in the extended vacuum module. Two of them are the stress tensors $T_{UV}$ and $\overline T_{IR}$ of the two factors in the folded theory. The third is an additional spin-two current, which we denote by
\begin{align}
  X=\phi^{\mathcal M_5}_{7/5}\,\bar\phi^{\mathcal M_4}_{3/5}\,, 
\end{align}
condensed in the vacuum. The theory $\mathcal T^{\,\prime}_{\mathcal B}$ further admits a $\mathbb Z_2$ automorphism that exchanges the two $\mathfrak{su}(2)_1$ factors in the numerator. In the folded theory, the RG domain wall becomes an RG boundary state, which can be represented as a rational twisted Cardy brane defined by this automorphism. The evaluation of a one-point correlator $\phi^{\mathcal M_5}\bar\phi^{\mathcal M_4}\in\mathcal T^{\,\prime}_{\mathcal B}$ in the presence of the RG brane is facilitated by passing to a fermionic description. As shown in \cite{Gaiotto:2012np}, $\mathcal T^{\,\prime}_{\mathcal B}$ contains an order-two simple current $J_\psi$ of conformal weight $1/2$. By fermionization with respect to $J_\psi$, we mean passing to the spin theory obtained by adjoining this simple current as a local fermion. This is a change of chiral-algebra description, rather than an additional RG flow:
\begin{align}
    \mathcal T^{\,\prime}_{\mathcal B}\ \xrightarrow[{\rm resp.\ to}\ \ J_\psi]{\rm fermionization}\ \widetilde{\mathcal T}^{\,\prime}_{\mathcal B}=\mathcal {SM}_{4,\,6}\times\mathcal T_{\psi}\,,
    \label{eq:fermionization-Bprime}
\end{align}
Here $\mathcal {SM}_{4,\,6}$ is the second $\mathcal N=1$ superconformal minimal model, while $\mathcal T_{\psi}$ is the theory of a free Majorana fermion. In this description, $J_\psi$ is represented by the Majorana field $\psi$, and the exchange of the two $\mathfrak{su}(2)_1$ numerator factors acts simply as $\psi\mapsto-\psi$. In the fermionic basis, the stress tensors of the UV and IR theories and $X$ are
\begin{align}
    &T_{UV}=\frac{2}{5}T_{\mathcal{SM}}-\frac{\sqrt 6}{5}G\,\psi+\frac{4}{5}T_{\psi}\notag\\
    &\overline T_{IR}=\frac{3}{5}T_{\mathcal{SM}}+\frac{\sqrt 6}{5}G\,\psi+\frac{1}{5}T_{\psi}\notag\\
    &X=\frac{2}{\sqrt 7}T_{\mathcal{SM}}-\frac{\sqrt 6}{2\sqrt 7}G\,\psi-\frac{4}{\sqrt 7}T_{\psi}\,,
\label{eq:N=1_basis}
\end{align}
Here $T_{\mathcal {SM}}$ and $G$ are the stress tensor and spin-$\frac{3}{2}$ supercurrent of $\mathcal S\mathcal M_{4,\,6}$, while $T_\psi$ and $\psi$ are the stress tensor and free fermion of $\mathcal T_\psi$. We normalize $X$ so that $X(z)\cdot X(w)\sim\frac{1}{(z-w)^4}+\cdots$.

Now we turn to the RG domain wall between $\mathcal P_3$ and $\mathcal M_{4}$. Using the folding trick, we can also embed the folded theory into a larger auxiliary one
\begin{align}
    \mathcal P_3\otimes \overline{\mathcal M_{4}}=\frac{\mathfrak{su}(2)_3}{\mathfrak{u}(1)_3}\times \frac{\mathfrak{su}(2)_2\times\mathfrak{su}(2)_1}{\mathfrak{su}(2)_3}
    \ \longrightarrow\ 
    \mathcal T_{\mathcal B}=\frac{\mathfrak{su}(2)_2\times\mathfrak{su}(2)_1}{\mathfrak{u}(1)_3}\,.
\end{align}
Recall that $\mathcal P_3$ can be obtained by gauging a spin-3 primary $W^{(3)}$ in $\mathcal M_{5}$. In the embedding of $\mathcal M_{5}\otimes \mathcal M_{4}\rightarrow \mathcal T_{\mathcal B}^{\,\prime}$, 
we have a spin-1 simple current $J$ in the chiral algebra $\mathcal B^{\,\prime}$ that can be decomposed in terms of the primaries in $\mathcal M_{5}\otimes \overline{\mathcal M_{4}}$ as
\begin{align}
    J=\left(\frac{2}{5}\right)_{\!\!\!\mathcal M_{5}}\!\!\!\!\otimes\left(\frac{3}{5}\right)_{\!\!\!\mathcal M_{4}}
    \oplus\ \
    \left(3\right)_{\!\mathcal M_{5}}\!\otimes\left(0\right)_{\!\mathcal M_{4}}
\end{align}
The folded theory $\mathcal T_B=\mathcal P_3\otimes \overline{\mathcal M_{4}}$ is then obtained by orbifolding the $\mathbb Z_2^J$ symmetry generated by $J$\,,
\begin{align}
\mathcal T_{\mathcal B}=\mathcal T_{\mathcal B}^{\,\prime}/\mathbb Z_2^J\,.
\end{align}
In $\mathcal T_{\mathcal B}$, as we have shown in Sec.~\ref{sec:sym}, 
\begin{align}
    J=\phi^{\mathcal P_3}_{2/5}\,\bar\phi^{\mathcal M_4}_{3/5}\equiv \phi^{UV}_{2/5}\,\bar\phi^{IR}_{3/5}\,,
\end{align}
is the spin-1 phantom current. Meanwhile, for the spin-2 phantom current $X$ in $\mathcal T^{\,\prime}_{\mathcal B}$, notice that the operator $\phi^{\mathcal M_5}_{7/5}$ is not primary in $\mathcal P_3$, but is instead a $W^{(3)}$ descendant of $\phi^{\mathcal P_3}_{2/5}$. Thus $X$ is a Virasoro primary but a descendant with respect to the chiral algebra $\mathcal B$, namely
\begin{align}
X=\phi^{\mathcal M_5}_{7/5}\,\bar\phi^{\mathcal M_4}_{3/5}=\left(W^{(3)}_{-1}\phi^{\mathcal P_3}_{2/5}\right)\bar\phi^{\mathcal M_4}_{3/5}\equiv \left(W^{(3)}_{-1}\phi^{UV}_{2/5}\right)\bar\phi^{IR}_{3/5}\,.
\end{align}
Besides the spin-2 operator $X$, as well as the two stress tensors, there are also two more spin-2 operators: the descendant of $J$, $\partial J$, and a new Virasoro (quasi-)primary $W^{-}$ \cite{Antinucci:2025uvj}, 
\begin{align}
   W^{-}\equiv -i\frac{1}{\sqrt 3}\phi^{UV}_{2/5}\,\partial\bar\phi^{IR}_{3/5}+i\frac{\sqrt 3}{2}\partial\phi^{UV}_{2/5}\,\bar\phi^{IR}_{3/5}\,\,,
   \label{eq:W_N=3}
\end{align}  
where we have normalized the OPE of $W^{-}(z)\cdot W^{-}(w) \sim \frac{1}{(z-w)^4}+\cdots$. One can compute the vacuum character of $\mathcal T_{\mathcal B}$, 
\begin{align}
    \chi_{\rm vac}^{\mathcal B}=q^{-\frac{\frac{4}{5}+\frac{7}{10}}{24}}\left(1+q+5q^2\cdots\right)\,,
\end{align}
and indeed finds that the coefficients in front of $q$ and $q^2$ match with our previous operator counting.

To establish the RG domain wall, we must further specify an automorphism that defines a twisted boundary condition for the RG brane in the folded picture. In the case of $\mathcal T_{\mathcal B}=\frac{\mathfrak{su}(2)_2\times\mathfrak{su}(2)_1}{\mathfrak{u}(1)_3}$, it is not clear how to identify such an automorphism. However, since the orbifolding $\mathcal T_{\mathcal B}=\mathcal T_{\mathcal B}^{\,\prime}/\mathbb Z_2^J$ leaves the two stress-tensor operators $T_{UV}$ and $T_{IR}$ intact, an explicit twist connecting them is transparent in the fermionic basis used above. Indeed, in $\mathcal T'_{\mathcal B}$, the mutual charges of $J$ and $J_\psi$ are both unity,
\begin{align}
    Q_{J}(J_\psi)=Q_{J_\psi}(J)=1
\end{align}
so the currents $J$ and $J_\psi$ will be maintained in $\mathcal {SM}_{4,\,6}\times\mathcal T_{\psi}$ and $\mathcal T_{\mathcal B}$ respectively. We summarize the above procedure in the following commutative diagram:
\begin{align}
\begin{tikzpicture}
\node at (-0.5,0.5) {$\mathcal T^{\,\prime}_{\mathcal B}$};
\node at (5.3,0.5) {$\mathcal {SM}_{4,\,6}\times\mathcal T_{\psi}$};
\node at (-0.5,-2.5) {$\mathcal T_{\mathcal B}$};
\node at (5.3,-2.5) {$\mathcal {SM}^D_{4,\,6}\times\mathcal T_{\psi}$};
\draw[->] (0,.5)--(4,.5);
\draw[->] (0,-2.5)--(4,-2.5);
\draw[->] (-.5,0)--(-.5,-2);
\draw[->] (5.3,0)--(5.3,-2);
\node at (2,.75) {fermionize};
\node at (2.1,0.25) {resp. to $J_\psi$};
\node at (2,-2.25) {fermionize};
\node at (2.1,-2.75) {resp. to $J_\psi$};
\node at (-1.2,-.8) {gauge};
\node at (-1.2,-1.3) {$\mathbb Z_2^J$};
\node at (6,-.8) {gauge};
\node at (6,-1.3) {$\mathbb Z_2^J$};
\end{tikzpicture}
\notag
\label{eq:CD}
\end{align}

It is useful to note that $\mathcal {SM}^D_{4,\,6}$, the $D$-type variant of $\mathcal {SM}_{4,\,6}$, coincides with the first $\mathcal N=2$ minimal model. In the vacuum module of an $\mathcal N=2$ superconformal algebra, we have a spin-1 $R$-current $J_R$, two spin-$\frac{3}{2}$ supercurrents $G_\pm$, and a spin-2 stress tensor $T_{\mathcal{SM}}$. Together with the vacuum module of $\mathcal T_\psi$, this gives the spin-1 current $J_R$ and five spin-two operators: $T_{\mathcal {SM}}$, $T_\psi$, $G_\pm\psi$, and $\partial J$. For later convenience, we introduce a basis adapted to the $\mathcal N=1$ superconformal subalgebra,
\begin{align}
    G=\frac{1}{\sqrt 2}(G_++G_-)\,,\quad {\rm and}\quad G_D=\frac{1}{\sqrt 2}(G_+-G_-)\,.
\end{align}
Clearly, the supercurrent $G$ is neutral under the $R$-current $J_R$. The fields $T_{\mathcal {SM}}$ and $G$ furnish a $\mathcal N=1$ superconformal sub-algebra. The detailed full $\mathcal N=2$ superconformal algebra OPEs are summarized in App.~\ref{app:OPE_N=3}. As we've mentioned, the vacuum module,
\begin{align}
\mathcal V_{\rm vac}^{\mathcal B'}\subset \mathcal V^{\mathcal N=1}_{\rm vac}\otimes \mathcal V^{\psi}_{\rm vac}\subset \mathcal V^{\mathcal N=2}_{\rm vac}\otimes \mathcal V^{\psi}_{\rm vac}\,, \quad{\rm and}\quad 
\mathcal V_{\rm vac}^{\mathcal B'}\subset \mathcal V_{\rm vac}^{\mathcal B}\subset \mathcal V^{\mathcal N=2}_{\rm vac}\otimes \mathcal V^{\psi}_{\rm vac}\,,
\label{eq:subset-diagram}
\end{align}
remains the same according to the orbifolding/fermionization commutative diagram \eqref{eq:CD}. Therefore, for the three operators $T_{UV}$, $T_{IR}$ and $X$, we still have the decomposition \eqref{eq:N=1_basis}. On the other hand, for the two extra operators, the spin-1 phantom current $J$ and spin-2 $W^-$ in $\mathcal V^{\mathcal B}_{\rm vac}$ can also be identified with the $\mathcal N=2$ superconformal algebra, and the vacuum module of free Majorana fermion $\mathcal T_\psi$. We summarize them all as below:
\begin{align}
&J=\phi^{UV}_{2/5}\,\bar\phi^{IR}_{3/5}=\sqrt 3 J_R\,,\notag\\[1ex]
&T_{UV}=\frac{2}{5}T_{\mathcal{SM}}-\frac{\sqrt 6}{5}G\,\psi+\frac{4}{5}T_{\psi}\notag\\[1ex]
&\overline T_{IR}=\frac{3}{5}T_{\mathcal{SM}}+\frac{\sqrt 6}{5}G\,\psi+\frac{1}{5}T_{\psi}\notag\\[1ex]
&X=\left(W^{(3)}_{-1}\phi^{UV}_{2/5}\right)\bar\phi^{IR}_{3/5}=\frac{2}{\sqrt 7}T_{\mathcal{SM}}-\frac{\sqrt 6}{2\sqrt 7}G\,\psi-\frac{4}{\sqrt 7}T_{\psi}\notag\\[1ex]
&W^{-}=-i\frac{1}{\sqrt 3}\phi^{UV}_{2/5}\,\partial\bar\phi^{IR}_{3/5}+i\frac{\sqrt 3}{2}\partial\phi^{UV}_{2/5}\,\bar\phi^{IR}_{3/5}
    =i\sqrt\frac{3}{2}G_D\psi
    \label{eq:N=2_basis}
\end{align}

In this fermionic basis, the action of the RG-brane twist on a product-theory state reduces to the corresponding twist overlap under $\psi\mapsto-\psi$. In App.~\ref{app:UV-IR-mixing}, we evaluate this overlap for a representative set of low-lying fields. The resulting disk one-point functions give explicit UV--IR mixing coefficients for the wall $\mathfrak D_3$, including a vanishing channel and a non-trivial overlap $C^{\mathrm{overlap}}=1/3$ in a two-dimensional branching space. For generic $N$, arbitrary one-point functions would require the presently unknown automorphism, or equivalently the full RG boundary state, and we leave this extension to future work.

\subsubsection*{Defect conformal manifold of the RG domain wall}
We now study the marginal deformation of the RG domain wall between $\mathcal P_3$ and $\mathcal M_{4}$, denoted by $\mathfrak D_3$. Since there exists a spin-1 phantom current $J$, as discussed in \cite{Antinucci:2025uvj}, the superconformal interface can be deformed by a one-dimensional local operator on the defect,
\begin{align}
    U_\theta=e^{i\theta Q_J}+h.c.\,,\quad{\rm with}\quad Q_J=\int_\gamma\frac{dz}{2\pi i}J(z)
\end{align}
The stress tensor $T_{UV}$ and $\overline T_{IR}$ will mix with the operator $W^-$. A hallmark feature of such a defect having conformal manifold is that its reflection/transmission rates have to depend on a moduli parameter. So in what follows, we will compute the transmission rate by employing the cluster decomposition equation \eqref{eq:cluster_decomp} \cite{Furuta:2025ahl}. 

First, it is worth noticing that $\mathfrak D_3$ can be obtained by fusing the Gaiotto wall of $\mathcal M_5$ to $\mathcal M_4$ to a topological interface induced from the $\mathbb Z_2^J$-orbifold. The transmission rate of $\mathfrak D_3$ should be the same as that of the Gaiotto wall, as a topological interface is always transparent. Therefore, only $T_{UV}$, $\overline T_{IR}$ and the operator $X$ can mix on the wall $\mathfrak D_3$. To make the origin of the reflection matrix explicit, let
\[
\boldsymbol v_{\mathrm{in}}
=\bigl(T_{UV},\overline T_{IR},X\bigr)^{\mathsf T},
\qquad
\boldsymbol v_{\mathrm{out}}
=\bigl(\overline T_{UV},T_{IR},\overline X\bigr)^{\mathsf T}.
\]
Let $A$ denote the $3\times3$ coefficient matrix multiplying $\bigl(T_{\mathcal{SM}},G\psi,T_\psi\bigr)^{\mathsf T}$ in the $T_{UV}$, $\overline T_{IR}$, and $X$ rows of \eqref{eq:N=2_basis}. In this basis the twist is $D=\operatorname{diag}(1,-1,1)$, because only $\psi$ changes sign. Transforming back therefore gives $M=ADA^{-1}$ and $\boldsymbol v_{\mathrm{out}}=M\boldsymbol v_{\mathrm{in}}$. Explicitly \cite{Furuta:2025ahl},
\begin{align}
    M=
    \begin{pmatrix}
        \frac{1}{5}\,, &\frac{32}{35}\,, &-\frac{4}{5\sqrt 7}\\[1ex]
        \frac{4}{5}\,, &\frac{3}{35}\,, &\frac{4}{5\sqrt 7}\\[1ex]
        -\frac{2}{\sqrt 7}\,, &\frac{16}{7\sqrt 7}\,, & \frac{5}{7}
    \end{pmatrix}.
    \label{eq:N3-reflection-matrix}
\end{align}
The matrix obeys
\[
M^2=\mathbf 1,
\qquad MGM^{\mathsf T}=G,
\qquad (1,1,0)M=(1,1,0),
\]
where
\[
G=\operatorname{diag}\!\left(\frac{c_{UV}}2,\frac{c_{IR}}2,1\right)
=\operatorname{diag}\!\left(\frac25,\frac7{20},1\right).
\]
These identities express, respectively, the order-two nature of the twist, preservation of the two-point-function Gram matrix, and conservation of the total stress tensor.

After applying the deformation by $U_\theta$ to $\mathfrak D_3$, we adopt the following ansatz for the reflections of $T_{UV}$ and $\overline T_{IR}$:
\begin{align}
& \overline T_{UV}=\alpha\,T_{UV}+\beta\, \overline T_{IR}+\gamma\,W^-+\delta\, X\notag\\[1ex]
& T_{IR}=(1-\alpha)\,T_{UV}+(1-\beta)\, \overline T_{IR}-\gamma\,W^--\delta\, X\,,
\label{eq:ansatz}
\end{align}
where we have imposed the energy conservation equation
\begin{align}
    \overline T_{UV}+T_{IR}=T_{UV}+\overline T_{IR}\,.
\end{align}
To further determine the mixing parameters $\{\alpha,\,\beta,\,\gamma,\,\delta\}$, we need the OPEs among $T_{UV}$, $\overline T_{IR}$, $W^{-}$, and $X$. In the fermionic basis \eqref{eq:N=2_basis}, they are readily computed and the results are collected in App.~\ref{app:OPE_N=3}. We then compute the OPEs $\overline T_{UV}(z)\cdot\overline T_{UV}(w)$, $T_{IR}(z)\cdot T_{IR}(w)$, and $\overline T_{UV}(z)\cdot T_{IR}(w)$. Cluster decomposition requires the last OPE to vanish,
\begin{align}
    \overline T_{UV}(z)\cdot T_{IR}(w)\sim 0\,,
    \label{eq:cluster_decomp_2}
\end{align}
while, for the first two, we need them to satisfy
\begin{align}
    &\overline T_{UV}(z)\cdot\overline T_{UV}(w)\sim\frac{\frac{2}{5}}{(z-w)^4}+\frac{2\overline T_{UV}(w)}{(z-w)^2}+\frac{\partial\overline T_{UV}(w)}{z-w}\notag\\[1ex]
    &T_{IR}(z)\cdot T_{IR}(w)\sim\frac{\frac{7}{20}}{(z-w)^4}+\frac{2T_{IR}(w)}{(z-w)^2}+\frac{\partial T_{IR}(w)}{z-w}\,.
    \label{eq:TT}
\end{align}
Eqs.~\eqref{eq:TT} are guaranteed once the most singular terms are satisfied, because the fields in the ansatz \eqref{eq:ansatz} are quasi-primary by definition. This gives
\begin{align}
    &\frac{2}{5}\alpha^2+\frac{7}{20}\beta^2+\gamma^2+\delta^2=\frac{2}{5}\notag\\[1ex]
    &\frac{2}{5}(1-\alpha)^2+\frac{7}{20}(1-\beta)^2+\gamma^2+\delta^2=\frac{7}{20}\,.
    \label{eq:TT_coef}
\end{align}
On the other hand, for eq.~\eqref{eq:cluster_decomp_2}, we collect the coefficients in front of all primaries,
\begin{align}
    &\frac{1}{(z-w)^4}:\qquad \frac{2}{5}\alpha(1-\alpha)+\frac{7}{20}\beta(1-\beta)-\gamma^2-\delta^2=0\notag\\[1ex]
    &\frac{J(w)}{(z-w)^3}:\qquad -\gamma\,\delta+\delta\,\gamma=0\notag\\[1ex]
    &\frac{T_{UV}(w)}{(z-w)^2}:\qquad 2\alpha(1-\alpha)-\frac{5}{2}\gamma^2-\frac{7}{2}\delta^2=0\notag\\[1ex]
    &\frac{T_{IR}(w)}{(z-w)^2}:\qquad 2\beta(1-\beta)-\frac{20}{7}\gamma^2-\frac{12}{7}\delta^2=0\notag\\[1ex]
    &\frac{X(w)}{(z-w)^2}:\qquad \frac{7}{5}(1-2\alpha)\delta+\frac{3}{5}(1-2\beta)\delta-\frac{1}{\sqrt{7}}\gamma^2+\frac{3}{\sqrt 7}\delta^2=0\notag\\[1ex]
    &\frac{W^{-}(w)}{(z-w)^2}:\qquad (1-2\alpha)\gamma+(1-2\beta)\gamma-\frac{1}{\sqrt{7}}\gamma\,\delta-\frac{1}{\sqrt 7}\delta\,\gamma=0
    \label{eq:CD_coef}
\end{align}
Solving eqs.~\eqref{eq:TT_coef} and \eqref{eq:CD_coef}, one can find a family of solutions parametrized by the coefficient $\gamma$,
\begin{align}
    \alpha=\frac{1}{5}\left(3\pm\sqrt{4-25\gamma^2}\,\right)\,,\quad
    \beta=\frac{8}{35}\left(2\mp\sqrt{4-25\gamma^2}\,\right)\,,\quad
    \delta=\frac{1}{5\sqrt 7}\left(-2\pm\sqrt{4-25\gamma^2\,}\right)\,.
\end{align}
Since the RG flow is between unitary CFTs, we would expect the coefficients $\alpha$ and $\beta$ real. Therefore, it's reasonable to re-parametrize $\gamma$ as
\begin{align}
    \gamma=\frac{2}{5}\sin\theta\,.
\end{align}
In terms of $\theta$, we have
\begin{align}
    \alpha=\frac{1}{5}(3-2\cos\theta)\,,\quad
    \beta=\frac{16}{35}(1+\cos\theta)\,,\quad
    \gamma=\frac{2}{5}\sin\theta\,,\quad
    \delta=-\frac{2}{5\sqrt 7}(1+\cos\theta)\,.
\end{align}
For $\theta = 0$, we simply recover the Gaiotto's RG domain wall as computed in \cite{Furuta:2025ahl}. To understand the physical properties of the defect conformal manifold, notice that the transmission rate $\mathcal{T}$ across the interface is determined by the cross two-point function $c_{UV\text{-}IR}$ between the stress tensors\cite{Quella:2006de, Brunner:2015vva, Meineri:2019ycm},
\begin{align}
    {T}_{IR}(z)\cdot{T}_{UV}(w)\sim\frac{c_{UV\text{-}IR}}{2(z-w)^4}\,.
\end{align}
From our ansatz \eqref{eq:ansatz}, taking the expectation value with ${T}_{IR}$ and $T_{UV}$ respectively, we obtain
\begin{align}
    c_{UV\text{-}IR}  = (1-\alpha)\,c_{UV} = \beta\,c_{IR}\,.
\end{align}
Therefore, the transmission rates in both directions are strictly proportional to $\beta$. One can find that, with $\theta$ increasing, $\beta$ decreases and the defect becomes less transparent. At $\theta = \pi$, we have
\begin{align}
    \alpha = 1 \,,\quad \beta = 0 \,,\quad \gamma = 0 \,,\quad \delta = 0 \,,
\end{align}
meaning the transmission rate strictly vanishes. The conformal defect becomes completely reflective, reducing to a trivially factorized interface.

\subsection{General case $N$}
We now proceed to the generic RG flow from $\mathcal P_N$ to $\mathcal M_{N+1}$. As we have shown in Sec.~\ref{sec:sym}, the deformation \eqref{eq:def} preserves various non-invertible symmetries. Among them, the line $\mathcal L_{[3,0]}$ in $\mathcal P_N$ penetrates the RG wall and becomes the line $\mathcal L_{[1,3]}$ in $\mathcal M_{N+1}$. Defect operators $\phi^{UV}_{[3,0]}$ and $\phi^{IR}_{[1,3]}$ are attached to the endpoints of these two lines, respectively. When the tensor product of the two theories is embedded into the auxiliary theory $\mathcal T_{\mathcal B}$,
\begin{align}
    \mathcal P_N\times\overline{\mathcal M_{N+1}}\longrightarrow\mathcal T_\mathcal B=\frac{\mathfrak {su}(2)_{N-1}\times\mathfrak{su}(2)_1}{\mathfrak{u}(1)_N}\,,
\end{align}
the tensor product of the two defect operators 
\begin{align}
    J=\phi^{UV}_{[3,0]}\,\bar\phi^{IR}_{[1,3]}\,,
\end{align}
as well as its descendants, condenses to the vacuum module of $\mathcal T_\mathcal B$ and thus becomes a local operator. For generic $N$, the vacuum module of $\mathcal T_\mathcal B$ is 
\begin{align}
    \chi_{\rm vac}^{\mathcal B}=\sum_{k=0}^{\left\lfloor N/2\right\rfloor}\chi^{UV}_{[2k+1,0]}\,\chi^{IR}_{[1,2k+1]}=q^{-\frac{c_{UV}+c_{IR}}{24}}\left(1+q+5q^2\cdots\right)\,,
\end{align}
As before, the operator $J$ corresponds to the linear $q$-term. For $N\geq3$, we collect all four Virasoro (quasi-)primaries of spin two. Besides $T_{UV}$ and $\overline T_{IR}$, there is a $W^{-}$ operator defined as in Eq.~\eqref{eq:W_N=3},
\begin{align}
    &W^{-}\equiv \frac{i}{\sqrt{2h_{UV}h_{IR}}}\left(h_{IR}\,\partial\phi^{UV}_{[3,0]}\,\bar\phi^{IR}_{[1,3]}-h_{UV}\,\phi^{UV}_{[3,0]}\,\partial\bar\phi^{IR}_{[1,3]}\right)\,,\notag\\[2ex]
    {\rm with}\quad &h_{UV}=\frac{2}{N+2}\,,\quad {\rm and}\quad h_{IR}=\frac{N}{N+2}\,.
\end{align}
In addition, as in the $\mathcal P_3$ case, the family of $\phi^{UV}_{[3,0]}$ contains a $W^{(3)}$ descendant that is nevertheless a Virasoro primary. For the neutral parafermion primary $[3,0]$, the $W^{(3)}_0$ eigenvalue vanishes. Equivalently, the $W^{(3)}$ charge is odd under the UV chiral charge-conjugation automorphism, while the label $s=0$ is fixed, and hence
\begin{align}
    W^{(3)}_0\,|[3,0]\rangle_{UV}=0\,.
\end{align}
We use this only as a representation-theoretic statement; the RG wall is not assumed to preserve this chiral automorphism factor by factor. Define
\begin{align}
    \phi^{UV}_{w}\equiv W^{(3)}_{-1}\phi^{UV}_{[3,0]}\,,
\end{align}
Since $W^{(3)}$ is a Virasoro primary of weight three, its modes obey
\[
[L_m^{UV},W_n^{(3)}]=(2m-n)W_{m+n}^{(3)}.
\]
Using $W_0^{(3)}|[3,0]\rangle_{UV}=0$ and the highest-weight conditions $W_{n>0}^{(3)}|[3,0]\rangle_{UV}=0$, we obtain
\begin{align}
    L_m^{UV}|w\rangle_{UV}
    &=(2m+1)W_{m-1}^{(3)}|[3,0]\rangle_{UV}=0,
    \qquad m\geq1\,.
    \label{eq:phi-w-Virasoro-primary}
\end{align}
For $m=1$ this follows from the vanishing $W_0^{(3)}$ charge, whereas for $m\geq2$ it follows from the highest-weight condition. Thus $\phi_w^{UV}$ is a Virasoro primary of weight $h_{UV}+1$. In $\mathcal T_\mathcal B$, the fourth Virasoro (quasi-)primary is therefore
\begin{align}
    X\equiv \phi^{UV}_w\,\bar\phi^{IR}_{[1,3]}\,.
\end{align}

\noindent 
We now need the OPEs among $T_{UV}$, $\overline T_{IR}$, $X$, and $W^{-}$ to determine the coefficients in the ansatz
\begin{align}
& \overline T_{UV}=\alpha\,T_{UV}+\beta\, \overline T_{IR}+\gamma\,W^-+\delta\, X\notag\\[1ex]
& T_{IR}=(1-\alpha)\,T_{UV}+(1-\beta)\, \overline T_{IR}-\gamma\,W^--\delta\, X\,,
\label{eq:ansatz2}
\end{align}

\paragraph{OPE at $\mathcal O\left((z-w)^{-3}\right)$ and $\mathcal O\left((z-w)^{-1}\right)$:}
The OPE of ansatz \eqref{eq:ansatz2} for odd power of $z-w$ trivially satisfies eq.~\eqref{eq:cluster_decomp}. First we claim that the OPE of $X(z)\cdot X(w)$ and $W^{-}(z)\cdot W^{-}(w)$ will not give non-trivial quasi-primaries at $\mathcal O\left((z-w)^{-3}\right)$ and $\mathcal O\left((z-w)^{-1}\right)$. Consider, for example, 
\begin{align}
    X(z)\cdot X(w)\sim \frac{1}{(z-w)^4}+\frac{O_3(w)}{(z-w)^3}+\frac{O_2(w)}{(z-w)^2}+\frac{O_1(w)}{z-w}\,.
\end{align}
The symmetry constraint $X(z)\cdot X(w)=X(w)\cdot X(z)$ simply implies that
\begin{align}
    O_3(w)=0\,,\quad {\rm and}\quad \mathcal O_1(w)=\frac{1}{2}\partial O_2(w)\,.
\end{align}
Therefore the diagonal terms at odd power of $z-w$ provide no constraints. On the other hand the cross terms in eq.~\eqref{eq:cluster_decomp}, e.g.
\begin{align}
    -\gamma\,\delta \ X(z)\cdot W^{-}(w)-\gamma\,\delta \ W^{-}(z)\cdot X(w)\bigg\vert_{(z-w)^{-1}}=\cdots-\gamma\,\delta\left(\frac{O_{prim}(w)}{z-w}+\frac{O_{prim}(z)}{w-z}\right)+\cdots\,,\notag
\end{align}
trivially satisfy the cluster decomposition principle, where $O_{prim}$ denotes any spin-three quasi-primary that may occur.

Finally, let us consider the cross terms between the stress tensors and $X$ or $W^{-}$ at $\mathcal O\left((z-w)^{-1}\right)$. Suppose that a spin-three quasi-primary $O_{prim}$ appears in the OPE between $T_{UV}$ and $X$,
\begin{align}
    T_{UV}(z)\cdot X(w)\sim \cdots+\frac{O_{prim}}{z-w}+\cdots\,,
\end{align}
one must have 
\begin{align}
    \overline T_{IR}(z)\cdot X(w)\sim \cdots -\frac{O_{prim}}{z-w}+\cdots\,,
\end{align}
because operator $X$ is a primary with respect to the total stress tensor $T=T_{UV}+\overline T_{IR}$ in the folded theory. Therefore the coefficient of $O_{prim}$ in $\overline T_{UV}(z)\cdot T_{IR}(w)$ is 
\begin{align}
\left(-\alpha\,\delta +\beta\,\delta\,\right)\,\frac{1}{z-w}+\left((1-\alpha)\,\delta-(1-\beta)\,\delta\,\right)\,\frac{1}{w-z}=0
\end{align}
Similar argument is applied to operator $W^{-}$ as well. Overall, we will have no any non-trivial constraints at $\mathcal O\left((z-w)^{-1}\right)$. 

\paragraph{OPE up to $\mathcal O\left((z-w)^{-2}\right)$:}
Therefore we only consider the OPE up to $\mathcal O\left((z-w)^{-2}\right)$. The OPE among $T_{UV}$ and $\overline T_{IR}$ are standard
\begin{align}
    &T_{UV}(z)\cdot T_{UV}(w)\sim \frac{c_{UV}/2}{(z-w)^4}+\frac{2T_{UV}(w)}{(z-w)^2}+\frac{\partial T_{UV}(w)}{z-w}\,,\quad
    {\rm with\ \ }c_{UV}=2\frac{N-1}{N+2}\notag\\[2ex]
    &T_{IR}(z)\cdot T_{IR}(w)\sim \frac{c_{IR}/2}{(z-w)^4}+\frac{2T_{IR}(w)}{(z-w)^2}+\frac{\partial T_{IR}(w)}{z-w}\,,\quad
    {\rm with\ \ }c_{IR}=1-\frac{6}{(N+1)(N+2)}\notag\\[2ex]
    &T_{UV}(z)\cdot \overline T_{IR}(w)\sim 0\,.
\end{align}
Because the operators factorize, the OPEs involving $X$ or $W^{-}$ with $T_{UV}$ or $T_{IR}$ follow directly from the respective constituent theories $\mathcal P_N$ and $\mathcal M_{N+1}$. Defining $\kappa\equiv i\sqrt{2h_{UV}h_{IR}}$, we find
\begin{align}
    &T_{UV}(z)\cdot W^{-}(w) \sim \frac{\kappa J(w)}{(z-w)^3}+\frac{\frac{\kappa}{2}\partial J(w)+W^{-}(w)}{(z-w)^2}\notag\\[2ex]
    &T_{IR}(z)\cdot W^{-}(w) \sim \frac{-\kappa J(w)}{(z-w)^3}+\frac{-\frac{\kappa}{2}\partial J(w)+W^{-}(w)}{(z-w)^2}\notag\\[2ex]
    &T_{UV}(z)\cdot X(w)\sim \frac{(h_{UV}+1)X(w)}{(z-w)^2}\notag\\[2ex]
    &T_{IR}(z)\cdot X(w)\sim \frac{h_{IR}\,X(w)}{(z-w)^2}
\end{align}
In addition, for the OPE among operators $X$ and $W^{-}$, notice that $\phi^{UV}_{[3,0]}$ and $\bar\phi^{IR}_{[1,3]}$ do not talk to each other, and their own fusion channels are the same as those of TDLs $\mathcal L_{[3,0]}^{UV}$ and $\mathcal L^{IR}_{[1,3]}$. Therefore we have
\begin{align}
    \left[\phi^{UV}_{[3,0]}\right]\cdot \left[\phi^{UV}_{[3,0]}\right]=\sum_{k=0}^{\min(2,N-2)}\left[\phi^{UV}_{[2k+1,0]}\right]\,,\qquad
    \left[\bar\phi^{IR}_{[1,3]}\right]\cdot \left[\bar \phi^{IR}_{[1,3]}\right]=\sum_{k=0}^{\min(2,N-2)}\left[\bar \phi^{IR}_{[1,2k+1]}\right]\,.
\end{align}
Recall that the operators $\phi^{UV}_{[2k+1,0]}$ and $\bar\phi^{IR}_{[1,2k+1]}$ have conformal weights
\begin{align}
    h^{UV}_{[2k+1,0]}=\frac{k(k+1)}{N+2}\,,\qquad 
    h^{IR}_{[1,2k+1]}=k^2-\frac{k(k+1)}{N+2}\,,
\end{align}
and all the operators $\phi^{UV}_{[2k+1,0]}\,\bar\phi^{IR}_{[1,2k+1]}$ reside in the vacuum module. Therefore, for the fusion channels of $W^{-}$ and $X$, we only need to collect terms admitting integer pole structures in the OPE and discard those with fractional branch cuts. Based on this observation, we have the following OPEs:
\begin{align}
    &X(z)\cdot X(w)\sim \frac{1}{(z-w)^4}+\frac{\frac{2(h_{UV}+1)}{c_{UV}}T_{UV}(w)+\frac{2h_{IR}}{c_{IR}}\overline T_{IR}(w)+C_{XXW^{-}}W^{-}(w)+C_{XXX}X(w)}{(z-w)^2}\notag\\[1ex]
    &X(z)\cdot W^{-}(w) \sim \frac{C_{JXW^{-}} J(w)}{(z-w)^3}+\frac{\frac{C_{JXW^{-}}}{2}\partial J(w)+C_{XW^{-}W^{-}}W^{-}(w)+C_{XXW^{-}}X(w)}{(z-w)^2}\notag\\[1ex] 
    &W^{-}(z)\cdot W^{-}(w) \sim \frac{1}{(z-w)^4}\notag\\[1ex]
    &\qquad\qquad\qquad\quad\ +\frac{\frac{2}{c_{UV}}T_{UV}(w)+\frac{2}{c_{IR}}\overline T_{IR}(w)+C_{XW^{-}W^{-}}X(w)+C_{W^{-}W^{-}W^{-}}W^{-}(w)}{(z-w)^2}\,,
\end{align}
where the OPE coefficients
\begin{align}
    &C_{XXW^{-}}=C_{W^{-}W^{-}W^{-}}=0\,,\quad C_{JXW^{-}}=i\sqrt{\frac{2N(N-2)}{(N-1)(N+4)}}\,,\notag\\[1ex]
    &C_{XXX}=\frac{3\sqrt2(N-4)}{\sqrt{(N-1)(N-2)(N+4)}}\,,\quad
    {\rm and}\quad C_{XW^{-}W^{-}}=\sqrt{\frac{2(N-2)}{(N-1)(N+4)}}\,.
\end{align}
The computation details on these OPE coefficients are summarized in the App.~\ref{app:OPE}.

\paragraph{One-parameter family of solutions:}
With these preparations, we can solve the OPE constraints for the mixing coefficients $\{\alpha,\,\beta,\,\gamma,\,\delta\}$. As before, the OPEs between the stress tensors give
\begin{align}
    &\frac{c_{UV}}{2}\alpha^2+\frac{c_{IR}}{2}\beta^2+\gamma^2+\delta^2=\frac{c_{UV}}{2}\notag\\[1ex]
    &\frac{c_{UV}}{2}(1-\alpha)^2+\frac{c_{IR}}{2}(1-\beta)^2+\gamma^2+\delta^2=\frac{c_{IR}}{2}\,,
\end{align}
and cluster decomposition \eqref{eq:cluster_decomp} gives:
\begin{align}
    &\frac{1}{(z-w)^4}:\qquad \frac{c_{UV}}{2}\,\alpha(1-\alpha)+\frac{c_{IR}}{2}\,\beta(1-\beta)-\gamma^2-\delta^2=0\notag\\[1ex]
    &\frac{J(w)}{(z-w)^3}:\qquad -\gamma\,\delta+\delta\,\gamma=0\notag\\[1ex]
    &\frac{T_{UV}(w)}{(z-w)^2}:\qquad 2\alpha(1-\alpha)-\frac{2}{c_{UV}}\gamma^2-\frac{2(h_{UV}+1)}{c_{UV}}\delta^2=0\notag\\[1ex]
    &\frac{T_{IR}(w)}{(z-w)^2}:\qquad 2\beta(1-\beta)-\frac{2}{c_{IR}}\gamma^2-\frac{2h_{IR}}{c_{IR}}\delta^2=0\notag\\[1ex]
    &\frac{X(w)}{(z-w)^2}:\qquad (h_{UV}+1)(1-2\alpha)\delta+h_{IR}(1-2\beta)\delta-C_{XW^{-}W^{-}}\,\gamma^2-C_{XXX}\delta^2=0\notag\\[1ex]
    &\frac{W^{-}(w)}{(z-w)^2}:\qquad (1-2\alpha)\gamma+(1-2\beta)\gamma-2C_{XW^{-}W^{-}}\,\gamma\,\delta=0
\end{align}
By parameterizing 
\begin{align}
    \gamma=\frac{\sqrt{2(N-1)}}{N+2}\sin\theta\,,
\end{align} 
we find
\begin{align}
    \alpha=\frac{N-2\cos\theta}{N+2}\,,\quad
    \beta=\frac{4(N+1)(1+\cos\theta)}{(N+4)(N+2)}\,,\quad
    \delta=-\frac{\sqrt{2(N-2)(N-1)}}{(N+2)\sqrt{N+4}}(1+\cos\theta)\,.
\end{align}
Therefore, overall, we compute the transmission rate of the RG domain wall $\mathfrak D_N(\theta)$ with defect conformal moduli parameter $\theta$,
\begin{align}
    \mathcal T_N(\theta)=\frac{2c_{UV\text{-}IR}}{c_{UV}+c_{IR}}=\frac{2\beta\,c_{IR}}{c_{UV}+c_{IR}}=\frac{8(N+1)(1+\cos\theta)}{3(N+2)^2}
\end{align}

As in the case $N=3$, the generic RG domain wall admits a modulus $\theta$ that interpolates between the RG wall and a fully reflecting, factorized conformal defect. Furthermore, as $N\rightarrow\infty$, $\beta\rightarrow0$ and the transmission rate $\mathcal T_N$ vanishes. This behavior highlights the genuinely non-perturbative nature of the RG flow from the $\mathbb{Z}_N$ parafermion theory to the minimal model $\mathcal{M}_{N+1}$. Unlike the flow between consecutive minimal models---where the central-charge difference $\Delta c\sim\mathcal O(1/N^3)$ is parametrically small and the fixed points are close in theory space, yielding a highly transparent domain wall---the parafermion flow is characterized by a macroscopic loss of degrees of freedom. As $N\to\infty$, the UV central charge approaches $c_{UV}=2$, corresponding geometrically to the $O(3)$ non-linear sigma model, while the IR central charge approaches $c_{IR}=1$, corresponding to a free compact boson. This $\Delta c\to1$ drop reflects a severe dynamical collapse of the target space, wherein an entire spatial dimension acquires an infinite mass and is integrated out. Algebraically, this macroscopic decoupling manifests within the extended chiral algebra $\mathcal T_{\mathcal B}$ of the folded theory: the structure constants governing the fusion of the phantom currents, such as $C_{XXX}$ and $C_{XW^-W^-}$, scale as $\mathcal O(1/\sqrt N)$. Consequently, in the large-$N$ limit, the couplings that mediate stress-tensor mixing vanish. Lacking these non-zero three-point couplings, local chiral energy fluctuations cannot be coherently transmitted along this parametrically long RG trajectory, forcing the RG domain wall to become completely reflective.

\section*{Acknowledgments}
We would like to thank Yuya Kusuki and Yunqin Zheng for useful discussions. J.C. is supported by the Fujian Provincial Natural Science Foundation of China (No.2025J01004), and the National Natural Science Foundation of China (Grants No.12247103). T.C. is supported by the Xiamen University ``Undergraduate Innovation Training Program" – Innovation Training (No.2025X1372).

\appendix
\section{Coset Construction}
\label{app:notation}

Coset CFT provides one of the most systematic constructions of RCFT. It realizes new chiral algebras by taking the commutant of a subalgebra inside a larger affine current algebra. Equivalently, starting from a Wess-Zumino-Witten theory with chiral algebra $\mathfrak g_k$, one gauges a chiral subalgebra $\mathfrak h_{k'}\subset \mathfrak g_k$. The resulting theory is denoted schematically by
\begin{align}
    \frac{\mathfrak g_k}{\mathfrak h_{k'}} ,
\end{align}
and its central charge is given by the difference
\begin{align}
    c_{\text{coset}}
    =
    c(\mathfrak g_k)-c(\mathfrak h_{k'}) .
\end{align}
For $\mathfrak{su}(2)_k$, we shall use
\begin{align}
    c\!\left(\mathfrak{su}(2)_k\right)
    =
    \frac{3k}{k+2}.
\end{align}
For a coset theory, the numerator and denominator factors are represented by Chern-Simons theories with opposite orientations:
\begin{align}
    \text{2D:}\quad 
    \frac{\mathfrak{su}(2)_{k_1}\times\cdots\times \mathfrak{su}(2)_{k_n}}
         {\mathfrak{su}(2)_{k'_1}\times\cdots\times \mathfrak{su}(2)_{k'_m}}
    \qquad
    \longleftrightarrow
    \qquad
    \text{3D:}\quad
    \prod_{i=1}^{n} \mathfrak{su}(2)_{k_i}
    \times
    \prod_{j=1}^{m} \mathfrak{su}(2)_{-k'_j}.
\end{align}
The negative levels in the denominator encode the fact that these degrees of freedom are gauged. Equivalently, their modular data appear complex conjugated in the coset modular matrices. Before imposing the coset constraints, a Wilson line is labeled by a collection of integrable representations,
\begin{align}
    \mathcal W_R
    =
    \mathcal W_{[\lambda_1,\ldots,\lambda_n;\,\mu_1,\ldots,\mu_m]},
\end{align}
where
\begin{align}
    \lambda_i=1,\ldots,k_i+1,
    \qquad
    \mu_j=1,\ldots,k'_j+1 .
\end{align}
Here we label $\mathfrak{su}(2)_k$ integrable representations by
\begin{align}
    a=1,\ldots,k+1,
    \qquad
    j=\frac{a-1}{2}.
\end{align}
For an abelian factor such as $\mathfrak{u}(1)_k$, we use charge labels
\begin{align}
    m\in \mathbb Z_{2k}.
\end{align}

The physical spectrum is obtained only after imposing the constraints generated by a distinguished topological line, which we denote by $\mathcal W_L$. This line is the \emph{identification line}. In the three-dimensional picture, it is an invertible Wilson line generating a discrete one-form symmetry. Its action has two effects.

First, fusing $\mathcal W_L$ with a Wilson line $\mathcal W_R$ maps the latter to another Wilson line with shifted representation labels. Since the identification line is topological, the two configurations are physically equivalent. This produces the field identification relation
\begin{align}
    R\sim L\times R .
\end{align}
Second, when $\mathcal W_L$ is braided around $\mathcal W_R$, the configuration acquires a monodromy phase
\begin{align}
    \mathcal M_{L,R}
    =
    \frac{S_{L,R}S_{0,0}}{S_{L,0}S_{0,R}} .
\end{align}
Only fields with trivial monodromy,
\begin{align}
    \mathcal M_{L,R}=1,
\end{align}
are mutually local with respect to the condensed line and therefore define admissible local operators in the coset theory. Thus, the familiar coset selection rules are reinterpreted as mutual-locality constraints with respect to the identification line.

A canonical example is the $\mathbb Z_N$ parafermion theory \eqref{eq:parafermion-coset}. Its three-dimensional realization is
\begin{align}
    \mathfrak{su}(2)_N \times \mathfrak{u}(1)_{-N}.
\end{align}
We label Wilson lines by pairs $\mathcal W_{[t,s]}$, where
\begin{align}
    t=1,\ldots,N+1,
    \qquad
    s\in \mathbb Z_{2N}.
\end{align}
Because the $\mathfrak{u}(1)$ factor appears with negative level in the coset, its contribution to the parafermion modular matrix is complex conjugated.

The identification line is
\begin{align}
    \mathcal W_L
    =
    \mathcal W_{[N+1,N]} .
\end{align}
Fusion with this line gives the field identification \eqref{eq:field-identification}. The monodromy of $\mathcal W_L$ around a general line $\mathcal W_{[t,s]}$ is
\begin{align}
    \mathcal M_{L,[t,s]}
    =
    (-1)^{t+s-1}.
\end{align}
Therefore, the mutual-locality condition is
\begin{align}
    \mathcal M_{L,[t,s]}=1
    \qquad
    \Longleftrightarrow
    \qquad
    t+s\in 2\mathbb Z+1 .
\end{align}
\eqref{eq:parafermion-S} is calculated by the modular matrices of the two factors
\begin{align}
    S^{\mathfrak{su}(2)_N}_{t,t'}
    =
    \sqrt{\frac{2}{N+2}}\,
    \sin\!\left(\frac{\pi tt'}{N+2}\right),
    \qquad
    S^{\mathfrak{u}(1)_N}_{s,s'}
    =
    \frac{1}{\sqrt{2N}}
    \exp\!\left(\frac{i\pi ss'}{N}\right),
    \label{eq:WZW-S-matrix}
\end{align}
Therefore,
\begin{align}
    S^{\mathcal P_N}_{[t,s],[t',s']}
    =
    2\,
    S^{\mathfrak{su}(2)_N}_{t,t'}\,
    \left(S^{\mathfrak{u}(1)_N}_{s,s'}\right)^*
    =
    \sqrt{\frac{4}{N(N+2)}}\,
    \sin\!\left(\frac{\pi tt'}{N+2}\right)
    \exp\!\left(-\frac{i\pi ss'}{N}\right).
\end{align}
The factor of $2$ accounts for the length of the generic simple-current orbit.

This construction also provides a useful way to construct auxiliary rational theories with extended chiral algebras. The auxiliary theory appearing in the folded description of the RG interface between consecutive minimal models provides an important example. For the product theory $\mathcal M_5\otimes \mathcal M_4$, it corresponds to the extended theory
\begin{align}
    \mathcal T_{\mathcal B}
    =
    \frac{\mathfrak{su}(2)_2\times \mathfrak{su}(2)_1\times \mathfrak{su}(2)_1}{\mathfrak{su}(2)_4}.
\end{align}
At the level of central charges, this gives
\begin{align}
    c(\mathcal T_{\mathcal B})
    =
    c(\mathfrak{su}(2)_2)+2c(\mathfrak{su}(2)_1)-c(\mathfrak{su}(2)_4)
    =
    \frac{3}{2}.
\end{align}
The Wilson lines of $\mathcal T_{\mathcal B}$ are labeled by four integers with identification
\begin{align}
    [t,r,\widetilde r,s]\sim
    [4-t,\;3-r,\;3-\widetilde r,\;6-s],
\end{align}
where
\begin{align}
    t=1,2,3,
    \qquad
    r,\widetilde r=1,2,
    \qquad
    s=1,\ldots,5 .
\end{align}
Here $t$ labels the $\mathfrak{su}(2)_2$ factor, $r$ and $\widetilde r$ label the two $\mathfrak{su}(2)_1$ factors, and $s$ labels the denominator $\mathfrak{su}(2)_4$ factor. 

The mutual-locality condition with respect to the identification line is
\begin{align}
    t+r+\widetilde r+s\in 2\mathbb Z .
\end{align}
Thus, the primary fields of $\mathcal T_{\mathcal B}$ are equivalence classes of labels $[t,r,\widetilde r,s]$ satisfying this parity constraint.

The conformal weight is computed from the coset formula
\begin{align}
    h_{\mathcal B}
    =
    h^{\mathfrak{su}(2)_2}_{t}
    +
    h^{\mathfrak{su}(2)_1}_{r}
    +
    h^{\mathfrak{su}(2)_1}_{\widetilde r}
    -
    h^{\mathfrak{su}(2)_4}_{s}
    \quad
    \mathrm{mod}\;1,
\end{align}
with
\begin{align}
    h^{\mathfrak{su}(2)_k}_{a}
    =
    \frac{a^2-1}{4(k+2)} .
\end{align}
Similarly, the modular matrix is obtained from the product of WZW modular matrices, with the denominator contribution complex conjugated. Since the $\mathfrak{su}(2)$ modular matrices are real, this gives
\begin{align}
    S^{\mathcal B}_{[t,r,\widetilde r,s],[t',r',\widetilde r',s']}
    &=
    2\,
    S^{\mathfrak{su}(2)_2}_{t,t'}\,
    S^{\mathfrak{su}(2)_1}_{r,r'}\,
    S^{\mathfrak{su}(2)_1}_{\widetilde r,\widetilde r'}\,
    S^{\mathfrak{su}(2)_4}_{s,s'}\notag\\[1ex]
    &=
    \sqrt{\frac{8}{27}}\,
    \sin\!\left(\frac{\pi tt'}{4}\right)
    \sin\!\left(\frac{\pi rr'}{3}\right)
    \sin\!\left(\frac{\pi \widetilde r\widetilde r'}{3}\right)
    \sin\!\left(\frac{\pi ss'}{6}\right).
\end{align}
This construction makes explicit how the extended theory $\mathcal T_{\mathcal B}$ inherits its rational data from the Chern--Simons description of the coset. In particular, its spectrum, conformal weights, field identifications, and modular transformations are all controlled by the same topological mechanism: the condensation of the identification line and the corresponding mutual-locality constraint.

\section{Fusion Rules in $\mathcal C_{\rm pres}$}
\label{app:sym}

In this appendix we spell out the fusion computation for the topological lines preserved by the parafermionic perturbation. The preserved lines found in the main text can be represented as
\begin{align}
    \mathcal L_{[t,0]},
    \qquad t\in 2\mathbb Z+1 .
\end{align}
Equivalently, writing
\begin{align}
    \ell=t-1 ,
\end{align}
the preserved labels are precisely the even-$\ell$ labels of $\mathfrak{su}(2)_N$.

We now compute their fusion rules. Let
\begin{align}
    i=\mathcal L_{[t_i,0]},
    \qquad
    j=\mathcal L_{[t_j,0]},
    \qquad
    t_i,t_j\in 2\mathbb Z+1,
\end{align}
while $k=\mathcal L_{[t_k,s_k]}$ is kept arbitrary. The Verlinde formula gives
\begin{align}
    N_{ij}^{\;\;k}
    =
    \sum_{h\in\mathcal P_N}
    \frac{
    S^{\mathcal P_N}_{ih}
    S^{\mathcal P_N}_{jh}
    \bigl(S^{\mathcal P_N}_{kh}\bigr)^{*}
    }{
    S^{\mathcal P_N}_{[1,0]h}
    } .
    \label{eq:app-verlinde}
\end{align}
Using \eqref{eq:parafermion-S}, we obtain
\begin{align}
    N_{ij}^{\;\;k}
    =
    \sum_{\mathcal L_{[t_h,s_h]}\in\mathcal P_N}
    \frac{4}{N(N+2)}
    \frac{
    \sin\!\left(\frac{\pi t_i t_h}{N+2}\right)
    \sin\!\left(\frac{\pi t_j t_h}{N+2}\right)
    \sin\!\left(\frac{\pi t_k t_h}{N+2}\right)
    }{
    \sin\!\left(\frac{\pi t_h}{N+2}\right)
    }
    \exp\!\left(\frac{\pi i s_k s_h}{N}\right).
    \label{eq:app-verlinde-expanded}
\end{align}

To evaluate the sum explicitly, we lift the sum over equivalence classes to a sum over the covering labels $\mathcal L_{[t_h,s_h]}$. The parity projection contributes $\frac12(1-(-1)^{t_h+s_h})$, and the quotient by the field identification contributes another factor of $\frac12$. Hence
\begin{align}
    N_{ij}^{\;\;k}
    =
    \frac{1}{N(N+2)}
    \sum_{t_h=1}^{N+1}
    \sum_{s_h=0}^{2N-1}
    \bigl(1-(-1)^{t_h+s_h}\bigr)
    F_{ijk}(t_h)
    \exp\!\left(\frac{\pi i s_k s_h}{N}\right),
    \label{eq:app-lifted-sum}
\end{align}
where
\begin{align}
    F_{ijk}(t_h)
    =
    \frac{
    \sin\!\left(\frac{\pi t_i t_h}{N+2}\right)
    \sin\!\left(\frac{\pi t_j t_h}{N+2}\right)
    \sin\!\left(\frac{\pi t_k t_h}{N+2}\right)
    }{
    \sin\!\left(\frac{\pi t_h}{N+2}\right)
    } .
\end{align}

The $s_h$-sum is a finite Fourier sum. We have
\begin{align}
    \sum_{s_h=0}^{2N-1}
    \exp\!\left(\frac{\pi i s_k s_h}{N}\right)
    =
    2N\,\delta^{(2N)}_{s_k,0},
\end{align}
and
\begin{align}
    \sum_{s_h=0}^{2N-1}
    (-1)^{s_h}
    \exp\!\left(\frac{\pi i s_k s_h}{N}\right)
    =
    2N\,\delta^{(2N)}_{s_k,N},
\end{align}
where $\delta^{(2N)}$ denotes the Kronecker delta modulo \(2N\). Therefore
\begin{align}
    \sum_{s_h=0}^{2N-1}
    \bigl(1-(-1)^{t_h+s_h}\bigr)
    \exp\!\left(\frac{\pi i s_k s_h}{N}\right)
    =2N\left(
    \delta^{(2N)}_{s_k,0}-(-1)^{t_h}\delta^{(2N)}_{s_k,N}
    \right).
\end{align}
Substituting this into \eqref{eq:app-lifted-sum} gives
\begin{align}
    N_{ij}^{\;\;k}=\delta^{(2N)}_{s_k,0}\,
    \mathcal N_{\ell_i,\ell_j}^{\;\;\ell_k}
    +\delta^{(2N)}_{s_k,N}\,\mathcal N_{\ell_i,\ell_j}^{\;\;N-\ell_k},
    \label{eq:app-selection-rule}
\end{align}
where
\begin{align}
    \ell_a=t_a-1,
    \qquad a=i,j,k,
\end{align}
and
\begin{align}
    \mathcal N_{\ell_i,\ell_j}^{\;\;\ell_k}
    =
    \frac{2}{N+2}
    \sum_{t_h=1}^{N+1}
    \frac{
    \sin\!\left(\frac{\pi(\ell_i+1)t_h}{N+2}\right)
    \sin\!\left(\frac{\pi(\ell_j+1)t_h}{N+2}\right)
    \sin\!\left(\frac{\pi(\ell_k+1)t_h}{N+2}\right)
    }{
    \sin\!\left(\frac{\pi t_h}{N+2}\right)
    } .
    \label{eq:app-su2-verlinde-sum}
\end{align}
In deriving the second term in \eqref{eq:app-selection-rule}, we used
\begin{align}
    (-1)^{t_h+1}
    \sin\!\left(\frac{\pi(\ell_k+1)t_h}{N+2}\right)
    =
    \sin\!\left(\frac{\pi(N-\ell_k+1)t_h}{N+2}\right).
\end{align}

The quantity $\mathcal N_{\ell_i,\ell_j}^{\;\;\ell_k}$ is precisely the $\mathfrak{su}(2)_N$ Verlinde coefficient. Equivalently,
\begin{align}
    \mathcal N_{\ell_i,\ell_j}^{\;\;\ell_k}
    =
    \sum_{\substack{
        \ell=
        |\ell_i-\ell_j|\\
        \mathrm{step}\;2
    }}^{
        \min(\ell_i+\ell_j,\;2N-\ell_i-\ell_j)
    }
    \delta_{\ell_k,\ell}.
    \label{eq:app-su2-fusion}
\end{align}
Thus the $s_h$-sum imposes the selection rule \eqref{eq:preserved-para}. The second possibility is not a new preserved object, because of \eqref{eq:TDL-id}. In terms of $\ell_k=t_k-1$, this identification sends
\begin{align}
    \ell_k\longmapsto N-\ell_k .
\end{align}
Therefore, after choosing representatives with $s=0$, the fusion rules of the preserved lines are
\begin{align}
    \mathcal L_{[\ell_i+1,0]}\times\mathcal L_{[\ell_j+1,0]}
    =\bigoplus_{\substack{
        \ell=
        |\ell_i-\ell_j|\\
        \mathrm{step}\;2
    }}^{
        \min(\ell_i+\ell_j,\;2N-\ell_i-\ell_j)
    }
    \mathcal L_{[\ell+1,0]},
    \qquad
    \ell_i,\ell_j,\ell\in 2\mathbb Z .
    \label{eq:app-preserved-fusion}
\end{align}
Equivalently, the fusion coefficients in the preserved sector are
\begin{align}
    N_{\ell_i,\ell_j}^{\;\;\ell_k}
    =
    \sum_{\substack{
        \ell=
        |\ell_i-\ell_j|\\
        \mathrm{step}\;2
    }}^{
        \min(\ell_i+\ell_j,\;2N-\ell_i-\ell_j)
    }
    \delta_{\ell_k,\ell},
    \qquad
    \ell_i,\ell_j,\ell_k\in 2\mathbb Z .
    \label{eq:app-SO3N-fusion}
\end{align}
These are exactly the fusion rules of the integer-spin subcategory of $\mathfrak{su}(2)_N$. Hence the preserved defect category is
\begin{align}
    \mathcal C_{\mathrm{pres}}
    \simeq
    \mathfrak{so}(3)_N .
\end{align}

\section{UV--IR Mixing Coefficients between $\mathcal P_3$ and $\mathcal M_{4}$}
\label{app:UV-IR-mixing}

In this appendix, we compute several disk one-point functions in the presence of the RG wall $\mathfrak D_3$ between $\mathcal P_3$ and $\mathcal M_4$. As in the explicit calculations for the Gaiotto wall, each one-point coefficient factorizes into a modular coefficient and a twist-overlap coefficient. The modular coefficient is fixed by the canonical interface in \eqref{eq:identity_interface} together with the Cardy coefficient of the auxiliary theory, whereas the overlap coefficient is determined after rewriting the relevant state in the fermionic basis
\begin{align}
    \widetilde{\mathcal T}_{\mathcal B}
    =\mathcal{SM}^{D}_{4,6}\times \mathcal T_\psi .
\end{align}
For the present example, the required $\mathbb Z_2$ automorphism reverses the Majorana fermion,
\begin{align}
    \Omega:\;\psi\;\longmapsto\;-\psi.
\end{align}
Accordingly, for a state $\ket{\phi}_{\widetilde{\mathcal B}}$ representing a product-theory operator, the normalized overlap factor is
\begin{align}
    C^{\mathrm{overlap}}=\frac{{}_{\widetilde {\mathcal B}}\bra{\phi}(-1)^{F_{\psi}}\ket{\phi}_{\widetilde {\mathcal B}}}
    {{}_{\widetilde {\mathcal B}}\langle\phi|\phi\rangle_{\widetilde {\mathcal B}}}.
    \label{eq:app-overlap-factor}
\end{align}
With the conventions of the main text, the disk one-point coefficient takes the form
\begin{align}
    \langle\bar\phi^{\mathcal M_4}_{[t,r]}\,\phi^{\mathcal P_3}_{[r,s]}\ket{\mathfrak D_3}=C^\mathrm{overlap}\frac{2^\frac{1}{4}\sqrt{S^{\mathfrak{su}(2)_2}_{1,t}S^{\mathfrak{u}(1)_3}_{0,s}}}{S^{\mathfrak{su}(2)_3}_{1,r}}.
    \label{eq:app-master-one-point}
\end{align}
The $\mathfrak{su}(2)_1$ label $d$ of the minimal-model field $[t,d,r]$ is suppressed here because the coset selection rule $t+d+r\in2\mathbb Z+1$ fixes it. The overall factor $2^{1/4}$ follows from the normalization of the orbifold topological interface; equivalently, it is fixed by requiring the vacuum-channel amplitude to equal the boundary entropy $g_{\mathfrak D_3}$ defined below.
When the corresponding branching space has multiplicity larger than one, the state $\ket{\phi}_{\widetilde{\mathcal B}}$ must first be identified by diagonalizing $L^{UV}_0$ within that space.

We first consider
\begin{align}
    \phi^{\mathcal P_3}_{[r,r-1]}\,
    \bar\phi^{\mathcal M_4}_{[r,r]}
    \subset \mathcal V^{\mathcal B}_{[r,1,r-1]},
    \qquad r=1,2,3.
\end{align}
The equality of conformal weights can be checked explicitly:
\begin{align}
    &h^{\mathcal B}_{[r,1,r-1]}
    -h^{\mathcal P_3}_{[r,r-1]}
    -h^{\mathcal M_4}_{[r,r]}
    \notag\\[1ex]
    &\quad=
    \left(\frac{r^2-1}{16}-\frac{(r-1)^2}{12}\right)
    -\left(\frac{r^2-1}{20}-\frac{(r-1)^2}{12}\right)
    -\frac{r^2-1}{80}
    =0.
\end{align}
For $r=1$ and $r=3$, the product-theory operators are the bottom components of their $\mathcal B$-modules. In the $\mathcal{SM}^{D}_{4,6}\times\mathcal T_\psi$ basis,
\begin{align}
    \ket{\phi^{\mathcal B}_{[r,1,r-1]}}
    \longleftrightarrow
    \begin{cases}
        \ket{\phi_0}_{\mathcal{SM}^{D}_{4,6}}
        \otimes\ket{0}_\psi,
        &r=1,\\[2ex]
        \ket{\phi_{\frac16,-\frac13}}_{\mathcal{SM}^{D}_{4,6}}
        \otimes\ket{0}_\psi,
        &r=3.
    \end{cases}
    \label{eq:app-bottom-components}
\end{align}
Here and below, $\phi_{\frac16}\equiv\phi_{\frac16,-\frac13}$ denotes the $h=\frac16$ primary of $\mathcal{SM}^{D}_{4,6}$ with $U(1)_R$ charge $q_R=-\frac13$; its charge-conjugate primary has $q_R=+\frac13$. Since neither state in \eqref{eq:app-bottom-components} contains a Majorana excitation, the twist acts trivially and $C^{\mathrm{overlap}}=1$.

The $r=2$ channel is different. Its total weight is
\begin{align}
    h^{\mathcal B}_{[2,1,1]}=\frac{5}{48}
    =\frac{1}{24}+\frac{1}{16},
\end{align}
so it lies in the Ramond sector of both factors in $\mathcal{SM}^{D}_{4,6}\times\mathcal T_\psi$. The chosen $\Omega$-twisted boundary state has no Ishibashi component in this Ramond--Ramond branching sector. It is therefore projected out of the overlap, and its overlap coefficient vanishes.

Using \eqref{eq:WZW-S-matrix},
\begin{align}
    S^{\mathfrak{su}(2)_2}_{1,1}
    =S^{\mathfrak{su}(2)_2}_{1,3}=\frac12,
    \qquad
    S^{\mathfrak{u}(1)_3}_{0,s}=\frac{1}{\sqrt6},
\end{align}
and substituting into \eqref{eq:app-master-one-point}, we obtain
\begin{align}
    \big\langle
    \bar\phi^{\mathcal M_4}_{[r,r]}\,
    \phi^{\mathcal P_3}_{[r,r-1]}
    \big|\mathfrak D_3\big\rangle
    =\begin{cases}
        \left[\dfrac{5}{6}\left(3+\sqrt5\right)\right]^{\frac14},
        &r=1,\\[2ex]
        0,
        &r=2,\\[2ex]
        \left[\dfrac{5}{6}\left(3-\sqrt5\right)\right]^{\frac14},
        &r=3.
    \end{cases}
    \label{eq:app-bottom-one-points}
\end{align}
The vanishing at $r=2$ is entirely due to the twist-overlap factor; the modular coefficient itself is nonzero.

We next consider the four operators obtained by setting $r=2$ in $\phi^{\mathcal P_3}_{[r,r-1\pm2]}\,\bar\phi^{\mathcal M_4}_{[r\pm1,r]}$. Their embeddings and conformal weights are
\begin{align}
    \phi^{\mathcal P_3}_{[2,5]}\,
    \bar\phi^{\mathcal M_4}_{[1,2]}
    &\subset \mathcal V^{\mathcal B}_{[1,2,5]},
    &h^{\mathcal P_3}+h^{\mathcal M_4}&=\frac{1}{15}+\frac{1}{10}=\frac16,
    \notag\\[1ex]
    \phi^{\mathcal P_3}_{[2,3]}\,
    \bar\phi^{\mathcal M_4}_{[1,2]}
    &\subset \mathcal V^{\mathcal B}_{[1,2,3]},
    &h^{\mathcal P_3}+h^{\mathcal M_4}&=\frac25+\frac{1}{10}=\frac12,
    \notag\\[1ex]
    \phi^{\mathcal P_3}_{[2,5]}\,
    \bar\phi^{\mathcal M_4}_{[3,2]}
    &\subset \mathcal V^{\mathcal B}_{[3,2,5]},
    &h^{\mathcal P_3}+h^{\mathcal M_4}&=\frac{1}{15}+\frac35=\frac23,
    \notag\\[1ex]
    \phi^{\mathcal P_3}_{[2,3]}\,
    \bar\phi^{\mathcal M_4}_{[3,2]}
    &\subset \mathcal V^{\mathcal B}_{[3,2,3]},
    &h^{\mathcal P_3}+h^{\mathcal M_4}&=\frac25+\frac35=1.
    \label{eq:app-four-embeddings}
\end{align}
The last operator is a level-one descendant of the $h^{\mathcal B}=0$ bottom component. In the fermionic basis, the four states are represented by
\begin{align}
    \ket{\phi^{\mathcal P_3}_{[2,5]}\,
    \bar\phi^{\mathcal M_4}_{[1,2]}}
    &\longleftrightarrow
    \ket{\phi_{\frac16}}_{\mathcal{SM}^{D}_{4,6}}
    \otimes\ket{0}_\psi,
    \notag\\[2ex]
    \ket{\phi^{\mathcal P_3}_{[2,3]}\,
    \bar\phi^{\mathcal M_4}_{[1,2]}}
    &\longleftrightarrow
    \ket{\phi_0}_{\mathcal{SM}^{D}_{4,6}}
    \otimes\psi_{-\frac12}\ket{0}_\psi,
    \notag\\[2ex]
    \ket{\phi^{\mathcal P_3}_{[2,5]}\,
    \bar\phi^{\mathcal M_4}_{[3,2]}}
    &\longleftrightarrow
    \mathrm{span}\!\left\{
    \ket{\phi_{\frac16}}_{\mathcal{SM}^{D}_{4,6}}
    \otimes\psi_{-\frac12}\ket{0}_\psi,
    \;
    G_{-\frac12}\ket{\phi_{\frac16}}_{\mathcal{SM}^{D}_{4,6}}
    \otimes\ket{0}_\psi
    \right\},
    \notag\\[2ex]
    \ket{\phi^{\mathcal P_3}_{[2,3]}\,
    \bar\phi^{\mathcal M_4}_{[3,2]}}
    &\longleftrightarrow
    \sqrt3\,J_{R,-1}\ket{\phi_0}_{\mathcal{SM}^{D}_{4,6}}
    \otimes\ket{0}_\psi.
    \label{eq:app-four-state-map}
\end{align}
The third channel has a two-dimensional branching space. Choose the basis
\begin{align}
    \ket{e_1}
    &=\ket{\phi_{\frac16}}_{\mathcal{SM}^{D}_{4,6}}
    \otimes\psi_{-\frac12}\ket{0}_\psi,
    \qquad
    \ket{e_2}
    =G_{-\frac12}\ket{\phi_{\frac16}}_{\mathcal{SM}^{D}_{4,6}}
    \otimes\ket{0}_\psi.
    \label{eq:app-e-basis}
\end{align}
The relevant mode algebra is
\begin{align}
    \{\psi_r,\psi_s\}=\delta_{r+s,0},
    \qquad
    \{G_r,G_s\}
    =2L^{\mathcal{SM}^{D}_{4,6}}_{r+s}
    +\frac{c}{3}\left(r^2-\frac14\right)\delta_{r+s,0},
    \qquad c=1.
\end{align}
It follows that
\begin{align}
    \langle e_1|e_1\rangle&=1,
    \qquad\langle e_1|e_2\rangle=0,
    \notag\\[1ex]
    \langle e_2|e_2\rangle
    &={}_{\mathcal{SM}^{D}_{4,6}}\!\bra{\phi_{\frac16}}
    G_{\frac12}G_{-\frac12}
    \ket{\phi_{\frac16}}_{\mathcal{SM}^{D}_{4,6}}
    =2h_{\phi_{\frac16}}
    =\frac13.
    \label{eq:level-1/2-basis}
\end{align}
Thus the basis is orthogonal but not orthonormal.

From \eqref{eq:N=2_basis},
\begin{align}
    &L^{\mathcal{SM}^D_{4,6}}_0\ket{e_1}=\frac{1}{6}\ket{e_1},\qquad  L^\psi_0\ket{e_1}=\frac{1}{2}\ket{e_1},\notag\\
    &L^{\mathcal{SM}^D_{4,6}}_0\ket{e_2}=\frac{2}{3}\ket{e_2},\qquad  L^\psi_0\ket{e_2}=0.
\end{align}
The mixed term acts as
\begin{align}
    &(G\psi)_0\ket{e_1}=G_{-\frac{1}{2}}\ket{\phi_\frac{1}{6}}_{\mathcal{SM}^D_{4,6}}\otimes \psi_{\frac{1}{2}}\psi_{-\frac{1}{2}}\ket 0_\psi=\ket{e_2},\notag\\[2ex]
    &(G\psi)_0\ket{e_2}=G_{\frac{1}{2}}G_{-\frac{1}{2}}\ket{\phi_\frac{1}{6}}_{\mathcal{SM}^D_{4,6}}\otimes \psi_{-\frac{1}{2}}\ket{0}_\psi=2L^{\mathcal{SM}^D_{4,6}}_0\ket{e_1}=\frac{1}{3}\ket{e_1}.
\end{align}
Substituting these relations into the decomposition of $T_{UV}$ in \eqref{eq:N=2_basis}, we find
\begin{align}
    L^{UV}_0\ket{e_1}=\frac{7}{15}\ket{e_1}-\frac{\sqrt{6}}{5}\ket{e_2},\quad L^{UV}_0\ket{e_2}=-\frac{\sqrt{6}}{15}\ket{e_1}+\frac{4}{15}\ket{e_2}
\end{align}
Using the convention that the $j$-th column contains the coefficients of $L^{UV}_0\ket{e_j}$, the matrix is therefore
\begin{align}
    (L^{UV}_0)_{(e_1,e_2)}=\begin{pmatrix}
        \frac{7}{15} & -\frac{\sqrt{6}}{15}\\[1ex]
        -\frac{\sqrt{6}}{5} & \frac{4}{15}
    \end{pmatrix}.
    \label{eq:app-LUV-matrix}
\end{align}
The matrix is not symmetric because the basis is not orthonormal. Nevertheless, it is Hermitian with respect to the Gram matrix $\operatorname{diag}(1,\frac13)$, as required.

Diagonalizing \eqref{eq:app-LUV-matrix} and normalizing with respect to \eqref{eq:level-1/2-basis} gives 
\begin{align}
    \ket{\mathcal O_{h^{UV}=\frac{1}{15}}}=\frac{1}{\sqrt{3}}\ket{e_1}+\sqrt{2}\ket{e_2},\quad\ket{\mathcal O_{h^{UV}=\frac{2}{3}}}=\sqrt{\frac{2}{3}}\ket{e_1}-\ket{e_2}.
    \label{eq:app-LUV-eigenstates}
\end{align}
Both states have total $\mathcal B$-weight $h^{\mathcal B}=\frac23$. Since $L^{\mathcal B}_0=L^{UV}_0+L^{IR}_0$ on the branching space, their IR weights are respectively
\begin{align}
    h^{IR}=\frac23-\frac{1}{15}=\frac35,
    \qquad
    h^{IR}=\frac23-\frac23=0.
\end{align}
Consequently, the first eigenstate in \eqref{eq:app-LUV-eigenstates} is the state associated with $\phi^{\mathcal P_3}_{[2,5]}\bar\phi^{\mathcal M_4}_{[3,2]}$, while the second is the companion branch associated, after field identification, with $\phi^{\mathcal P_3}_{[1,2]}\bar{\phi}_{[1,1]}^{\mathcal M_4}$.

The twist acts diagonally on the basis,
\begin{align}
    \Omega\ket{e_1}=-\ket{e_1},\qquad\Omega\ket{e_2}=\ket{e_2},
\end{align}
Therefore,
\begin{align}
    C^\mathrm{overlap}_{\phi^{\mathcal P_3}_{[2,5]}\,\bar\phi^{\mathcal M_4}_{[3,2]}}=-\left|\frac{1}{\sqrt3}\right|^2+\frac{2}{3}=\frac{1}{3}.
\end{align}
For the remaining one-dimensional channels, the overlap is determined directly by the parity of the Majorana excitation:
\begin{align}
    C^\mathrm{overlap}_{\phi^{\mathcal P_3}_{[2,5]}\,\bar\phi^{\mathcal M_4}_{[1,2]}}=1,\qquad
    C^\mathrm{overlap}_{\phi^{\mathcal P_3}_{[2,3]}\,\bar\phi^{\mathcal M_4}_{[1,2]}}=-1,\qquad
    C^\mathrm{overlap}_{\phi^{\mathcal P_3}_{[2,3]}\,\bar\phi^{\mathcal M_4}_{[3,2]}}=1.
\end{align}

For all four channels, the modular factor is the same:
\begin{align}
    \frac{2^{\frac14}
    \sqrt{S^{\mathfrak{su}(2)_2}_{1,1}
    S^{\mathfrak{u}(1)_3}_{0,5}}}
    {S^{\mathfrak{su}(2)_3}_{1,2}}
    =
    \frac{2^{\frac14}
    \sqrt{S^{\mathfrak{su}(2)_2}_{1,3}
    S^{\mathfrak{u}(1)_3}_{0,3}}}
    {S^{\mathfrak{su}(2)_3}_{1,2}}
    =\left[\frac56\left(3-\sqrt5\right)\right]^{\frac14}.
    \label{eq:app-common-modular-factor}
\end{align}

Finally, we obtain
\begin{align}
    &\langle\bar\phi^{\mathcal M_4}_{[1,2]}\,\phi^{\mathcal P_3}_{[2,5]}\ket{\mathfrak D_3}=\big(\frac{5}{6}(3-\sqrt{5})\big)^\frac{1}{4},\qquad
    \langle\bar\phi^{\mathcal M_4}_{[1,2]}\,\phi^{\mathcal P_3}_{[2,3]}\ket{\mathfrak D_3}=-\big(\frac{5}{6}(3-\sqrt{5})\big)^\frac{1}{4},\notag\\[2ex]
    &\langle\bar\phi^{\mathcal M_4}_{[3,2]}\,\phi^{\mathcal P_3}_{[2,5]}\ket{\mathfrak D_3}=\frac{1}{3}\big(\frac{5}{6}(3-\sqrt{5})\big)^\frac{1}{4},\qquad
    \langle\bar\phi^{\mathcal M_4}_{[3,2]}\,\phi^{\mathcal P_3}_{[2,3]}\ket{\mathfrak D_3}=\big(\frac{5}{6}(3-\sqrt{5})\big)^\frac{1}{4}.
\end{align}
It is useful to summarize these amplitudes after dividing by the vacuum overlap
\begin{align}
    g_{\mathfrak D_3}
    \equiv\langle0|\mathfrak D_3\rangle
    =\left[\frac56\left(3+\sqrt5\right)\right]^{\frac14},
    \qquad
    \varphi\equiv\frac{1+\sqrt5}{2}.
\end{align}
With unit two-point normalization and the phases fixed by the state identifications above, the normalized $r=1,2,3$ bottom-component amplitudes are
\begin{align}
    \widehat{\mathcal M}_{\mathrm{bottom}}
    =\left(1,0,\varphi^{-1}\right).
\end{align}
For the remaining four channels, order the IR rows as $([1,2],[3,2])$ and the UV columns as $([2,5],[2,3])$. The corresponding normalized mixing block is
\begin{align}
    \widehat{\mathcal M}
    =\varphi^{-1}
    \begin{pmatrix}
        1&-1\\[1ex]
        \frac13&1
    \end{pmatrix}.
\end{align}
The relative minus sign is phase-convention dependent and changes if the corresponding UV field is rephased.

\section{OPE in the $\mathcal {SM}^D_{4,\,6}\times\mathcal T_{\psi}$-basis}
\label{app:OPE_N=3}
In this appendix, we collect the details of the $\mathcal N=2$ superconformal algebra, and the OPE computation of cluster decomposition equation~\eqref{eq:cluster_decomp} in the basis of $\mathcal {SM}^D_{4,\,6}\times\mathcal T_{\psi}$.

Recall that we introduce an $\mathcal N=2$ basis consistent with $\mathcal N=1$ superconformal algebra,
\begin{align}
    G=\frac{1}{\sqrt 2}(G_++G_-)\,,\quad {\rm and}\quad G_D=\frac{1}{\sqrt 2}(G_+-G_-)\,.
\end{align}
In the $(G,\, G_D)$-basis, we rewrite the standard $\mathcal N=2$ superconformal algebra OPE as
\begin{align}
    &T_{\mathcal {SM}}(z)\cdot T_{\mathcal {SM}}(w) \sim \frac{1/2}{(z-w)^4} + \frac{2T_{\mathcal {SM}}(w)}{(z-w)^2} + \frac{\partial T_{\mathcal {SM}}(w)}{z-w} \notag\\[1ex]
    &T_{\mathcal {SM}}(z)\cdot J_R(w) \sim \frac{J_R(w)}{(z-w)^2} + \frac{\partial J_R(w)}{z-w} \notag\\[1ex]
    &J_R(z)\cdot J_R(w) \sim \frac{1/3}{(z-w)^2} \notag\\[1ex]
    &T_{\mathcal {SM}}(z)\cdot G(w) \sim \frac{\frac{3}{2}G(w)}{(z-w)^2} + \frac{\partial G(w)}{z-w}\,,\quad 
    T_{\mathcal {SM}}(z)\cdot G_D(w) \sim \frac{\frac{3}{2}G_D(w)}{(z-w)^2} + \frac{\partial G_D(w)}{z-w}\notag\\[1ex]
    &J_R(z)\cdot G(w) \sim \frac{G_D(w)}{z-w}\,,\quad
    J_R(z)\cdot G_D(w) \sim \frac{G(w)}{z-w}\notag\\[1ex]
    &G(z)\cdot G(w) \sim \frac{2/3}{(z-w)^3}  + \frac{2T_{\mathcal {SM}}(w) }{z-w}\,,\quad
    G_D(z)\cdot G_D(w) \sim -\frac{2/3}{(z-w)^3}  - \frac{2T_{\mathcal {SM}}(w) }{z-w}\,,\quad\notag\\[1ex]
    &G(z)\cdot G_D(w) \sim - \frac{2J_R(w)}{(z-w)^2} - \frac{ \partial J_R(w)}{z-w}\,,
    \label{eq:N=2_OPE}
\end{align}
where we have put the central charge $c=1$ of $\mathcal {SM}_{4,\,6}^D$.

In the fermionic basis, we can spell all these spin-2 quasi-primaries as in eq.\eqref{eq:N=2_basis}. We compute their OPEs from eq.~\eqref{eq:N=2_OPE}, and summarize below:
\begin{align}
    &T_{UV}(z)\cdot T_{UV}(w)\sim \frac{\frac{2}{5}}{(z-w)^4}+\frac{2T_{UV}(w)}{(z-w)^2}+\frac{\partial T_{UV}(w)}{z-w}\notag\\[1ex]
    &T_{UV}(z)\cdot \overline T_{IR}(w)\sim 0\notag\\[1ex]
    &T_{UV}(z)\cdot X(w) \sim \frac{\frac{7}{5}X(w)}{(z-w)^2}\notag\\[1ex]
    &T_{UV}(z)\cdot W^{-}(w) \sim \frac{\frac{2\sqrt 3\,i}{5}J(w)}{(z-w)^3}+\frac{\frac{\sqrt 3\,i}{5}\partial J(w)+W^{-}(w)}{(z-w)^2}\notag\\[1ex]
    &\overline T_{IR}(z)\cdot \overline T_{IR}(w)\sim \frac{\frac{7}{20}}{(z-w)^4}+\frac{2\overline T_{IR}(w)}{(z-w)^2}+\frac{\partial \overline T_{IR}(w)}{z-w}\notag\\[1ex]
    &\overline T_{IR}(z)\cdot X(w) \sim \frac{\frac{3}{5}X(w)}{(z-w)^2}\notag\\[1ex]
    &\overline T_{IR}(z)\cdot W^{-}(w)\sim \frac{-\frac{2\sqrt 3\,i}{5}J(w)}{(z-w)^3}+\frac{-\frac{\sqrt 3\,i}{5}\partial J(w)+W^{-}(w)}{(z-w)^2}\notag\\[1ex]
    &X(z)\cdot X(w)\sim \frac{1}{(z-w)^4}+\frac{\frac{7}{2}T_{UV}(w)+\frac{12}{7}\overline T_{IR}(w)-\frac{3}{\sqrt 7}X(w)}{(z-w)^2}\notag\\[1ex]
    &X(z)\cdot W^{-}(w) \sim \frac{\sqrt{\frac{3}{7}}\,i J(w)}{(z-w)^3}+\frac{\sqrt{\frac{3}{7}}\frac{i}{2}\partial J(w)+\frac{1}{\sqrt{7}}W^{-}(w)}{(z-w)^2}\notag\\[1ex]  
    &W^{-}(z)\cdot W^{-}(w) \sim \frac{1}{(z-w)^4}+\frac{\frac{5}{2}T_{UV}(w)+\frac{20}{7}\overline T_{IR}(w)+\frac{1}{\sqrt 7}X(w)}{(z-w)^2}\,.
\end{align}
Here we only collect the OPE terms up to $\mathcal O((z-w)^{-2})$ because the $(z-w)^{-1}$ order contains spin-3 operators that are irrelevant to our discussion.

\section{OPE for Operators $J$, $X$ and $W^{-}$}
\label{app:OPE}

In this appendix, we provide the detailed computation of the OPE coefficients involving the phantom currents and their descendants. The relevant operators in the UV and IR theories are respectively
\begin{align}
    \phi^{UV}\equiv\phi^{\mathcal P_N}_{[3,0]},
    \qquad
    \phi^{IR}\equiv\phi^{\mathcal M_{N+1}}_{[1,3]} .
\end{align}
In the folded picture, the composite operators of interest are
\begin{align}
    X=\phi_{w}^{UV} \bar\phi^{IR},
    \qquad
    \phi_{w}^{UV}\equiv W^{(3)}_{-1}\phi^{UV} ,
\end{align}
and
\begin{align}
    W^-=\frac{i}{\sqrt{2h^{UV}h^{IR}}}
    \left(
    h^{IR}\partial\phi^{UV}\bar\phi^{IR}
    -h^{UV}\phi^{UV}\partial\bar\phi^{IR}
    \right), \quad J=\phi^{UV}\bar\phi^{IR} .
\end{align}
We shall first determine the relevant three-point coefficients in the $\mathcal P_N$ sector, and then combine them with the corresponding minimal-model contribution.

The fundamental strategy is to lift the computation from the parafermion theory to the parent $\mathfrak{su}(2)_N$ WZW model. Under the coset decomposition, a primary field $\Phi_{j,m}$ in the $\mathfrak{su}(2)_N$ theory factorizes into a parafermion primary $\phi_{[t,s]}^{\mathcal{P}_N}$ and a free compact boson vertex operator from the $\mathfrak{u}(1)_N$ sector. Consequently, the OPE coefficients in $\mathcal{P}_N$ can be systematically extracted by computing the corresponding correlators in the $\mathfrak{su}(2)_N$ WZW model, which are exactly known in terms of Wigner $3j$-symbols and Knizhnik-Zamolodchikov (KZ) structure constants\cite{Knizhnik:1984nr}, and subsequently stripping off the trivial $\mathfrak{u}(1)_N$ free-boson contributions. Specifically, let's first focus on the operator
\begin{align}
    \phi^{UV} \equiv \phi^{\mathcal{P}_N}_{[3,0]}\,.
\end{align}
The affine $\mathfrak{su}(2)_N$ representation decomposes as
\begin{align}
    \mathcal H^{\mathfrak{su}(2)_N}_{t-1}=\bigoplus_{\substack{s=0,\dots,2N-1\\ t+s\ {\rm odd}}}
    \mathcal H^{\mathcal P_N}_{[t,s]}
    \otimes
    \mathcal H^{\mathfrak u(1)_N}_{s}.
\end{align}
Under this decomposition, the WZW quantum numbers are
\begin{align}
    j=\frac{t-1}{2},\qquad m=\frac{s}{2}.
\end{align}
In particular,
\begin{align}
    \phi^{UV}
    \quad\longleftrightarrow\quad
    \Phi_{j=1,m=0}
\end{align}
in the $\mathfrak{su}(2)_N$ theory. Moreover, the conformal weights satisfy
\begin{align}
    h^{UV}+h^{IR}=\frac{2}{N+2}+\frac{N}{N+2}=1 .
    \label{eq:spin-1}
\end{align}
We now construct the level-one descendant corresponding to $\phi_{w}^{UV}=W^{(3)}_{-1}\phi^{UV}$. A natural basis for the relevant level-one subspace is
\begin{align}
    v_+ = J^+_{-1}\Phi_{1,-1},\qquad
    v_0 = J^0_{-1}\Phi_{1,0},\qquad
    v_- = J^-_{-1}\Phi_{1,1} .
\end{align}
In the following, we suppress this common spin label and use the shorthand
\begin{align}
    \Phi_m \equiv \Phi_{1,m}.
\end{align}
Thus all fields $\Phi_m$ appearing below are understood to belong to the spin-one representation of $\mathfrak{su}(2)_N$. We therefore write
\begin{align}
    |\tilde \phi_{w}^{UV}\rangle=a v_+ + b v_0 + c v_- .
\end{align}
The state $|\tilde \phi_{w}^{UV}\rangle$ is fixed by two conditions.

First, it should be annihilated by the positive $\mathfrak u(1)$ mode,
\begin{align}
    J^0_1|\tilde \phi_{w}^{UV}\rangle=0 .
\end{align}
Using
\begin{gather}
    [J^+_m,J^-_n]=2J^0_{m+n}+Nm\delta_{m+n,0},\qquad
    [J^0_m,J^\pm_n]=\pm J^\pm_{m+n},\notag\\[0.5ex]
    [J^0_m,J^0_n]=\frac{N}{2}m\delta_{m+n,0},
    \label{eq:A1-alg}
\end{gather}
this condition gives
\begin{align}
    \sqrt 2\,a+\frac{N}{2}b-\sqrt 2\,c=0 .
\end{align}
Second, $|\tilde \phi_{w}^{UV}\rangle$ should be quasi-primary in the
$\mathcal P_N$ theory,
\begin{align}
    L^{\mathcal P_N}_1|\tilde \phi_{w}^{UV}\rangle=0 .
\end{align}
This gives
\begin{align}
    a+c=0 .
\end{align}
Solving these two constraints, we may choose the representative
\begin{align}
    |\tilde \phi_{w}^{UV}\rangle=-J^+_{-1}\Phi_{-1}+J^-_{-1}\Phi_{1}+\frac{4\sqrt 2}{N}J^0_{-1}\Phi_{0} .
\end{align}
After normalization, the state is
\begin{align}
    |\phi_{w}^{UV}\rangle=\sqrt{\frac{N}{2(N-2)(N+4)}}\left(-J^+_{-1}\Phi_{-1}+J^-_{-1}\Phi_{1}+\frac{4\sqrt 2}{N}J^0_{-1}\Phi_{0}\right).
    \label{eq:D-state}
\end{align}
We next evaluate the required three-point coefficients. The chiral three-point function in the $\mathfrak{su}(2)_N$ WZW theory takes the form \cite{Knizhnik:1984nr,Fateev:1985mm}
\begin{align}
    \big\langle
    \Phi_{j_1,m_1}(z_1)
    \Phi_{j_2,m_2}(z_2)
    \Phi_{j_3,m_3}(z_3)
    \big\rangle
    &=
    C
    \begin{pmatrix}
        j_1 & m_1\\
        j_2 & m_2\\
        j_3 & m_3
    \end{pmatrix}
    z_{12}^{D_{j_3}-D_{j_1}-D_{j_2}}z_{13}^{D_{j_2}-D_{j_3}-D_{j_1}}z_{23}^{D_{j_1}-D_{j_2}-D_{j_3}} ,
\end{align}
where
\begin{align}
    C
    \begin{pmatrix}
        j_1 & m_1\\
        j_2 & m_2\\
        j_3 & m_3
    \end{pmatrix}
    =
    K(j_1,j_2,j_3)
    \begin{pmatrix}
        j_1 & j_2 & j_3\\
        m_1 & m_2 & m_3
    \end{pmatrix}.
\end{align}
Here the second factor is the Wigner $3j$-symbol, while $K(j_1,j_2,j_3)$ contains the remaining $N$-dependent structure constant. Since only the case $j_1=j_2=j_3=1$ is needed below, we denote
\begin{align}
    K_N \equiv K(1,1,1).
\end{align}
The affine currents are expanded as
\begin{align}
    J^a(z)=\sum_{n\in\mathbb Z}J^a_n z^{-n-1},
\end{align}
Equivalently, if the current is expanded in a local coordinate around an insertion point $z_i$, the modes are defined by
\begin{align}
    J^a_n\mathcal O_i(z_i)=\oint_{z_i}\frac{dz}{2\pi i}(z-z_i)^nJ^a(z)\mathcal O_i(z_i).
    \label{eq:local-insertion}
\end{align}
We fix the three insertion points by global conformal symmetry as
\begin{align}
    z_1=\infty,\qquad z_2=1,\qquad z_3=0,
\end{align}
and use the shorthand
\begin{align}
    \big\langle
    \mathcal O_i(\infty)\mathcal O_j(1)\mathcal O_k(0)
    \big\rangle:=\lim_{z_1\to\infty}z_1^{2h_1}\big\langle\mathcal O_i(z_1)\mathcal O_j(1)\mathcal O_k(0)\big\rangle .
\end{align}
In the applications below, the fields are affine primaries $\Phi_m$ and their level-one descendants $v_i$. Thus the positive modes truncate as
\begin{align}
    J^a_{n\geq2}v_i=0,\quad J^a_{n\geq1}\Phi_m=0.
    \label{eq:state-condition}
\end{align}
Consequently, in the contour manipulations below only the $J^a_0$ and $J^a_1$ terms can contribute.

Since $J^a(z)$ is a meromorphic current, a contour encircling one insertion can be deformed to contours encircling the other insertions. We first consider
\begin{align}
    \big\langle\mathcal O_i(\infty)\mathcal O_j(1)(J^a_{-1}\mathcal O_k)(0)\big\rangle.
\end{align}
By the local definition \eqref{eq:local-insertion},
\begin{align}
    J^a_{-1}\mathcal O_k(0)=\oint_0\frac{dz}{2\pi i}\frac{1}{z}J^a(z)\mathcal O_k(0).
\end{align}
Hence
\begin{align}
    \big\langle\mathcal O_i(\infty)\mathcal O_j(1)(J^a_{-1}\mathcal O_k)(0)\big\rangle=\oint_0\frac{dz}{2\pi i}\frac{1}{z}\big\langle\mathcal O_i(\infty)\mathcal O_j(1)J^a(z)\mathcal O_k(0)\big\rangle.
\end{align}
We now deform the contour around $z=0$ to contours around $z=1$ and
$z=\infty$. Near $z=1$,
\begin{align}
    \frac{1}{z}=\frac{1}{1+(z-1)}=1-(z-1)+(z-1)^2-\cdots .
\end{align}
Therefore,
\begin{align}
    \oint_{1}\frac{dz}{2\pi i}\,
    \frac{1}{z}J^a(z)
    =
    J^a_0-J^a_1+J^a_2-\cdots .
\end{align}
After contour deformation, this contribution comes with an overall minus sign. Using \eqref{eq:state-condition}, it reduces to
\begin{align}
    -J^a_0+J^a_1 .
\end{align}
For the contribution from infinity, we introduce the local coordinate
\begin{align}
    z=\frac{1}{w}.
\end{align}
Since $J^a$ is a weight-one current,
\begin{align}
    J^a(z)\,dz=J^a(w)\,dw .
\end{align}
Thus
\begin{align}
    \oint_{\infty}\frac{dz}{2\pi i}\,
    \frac{1}{z}J^a(z)
    =-\oint_{w=0}\frac{dw}{2\pi i}\,w J^a(w)
    =-J^a_1 .
\end{align}
The minus sign from this change of coordinate is canceled by the minus sign from the contour deformation. We obtain
\begin{align}
    \big\langle\mathcal O_i(\infty)\mathcal O_j(1)(J^a_{-1}\mathcal O_k)(0)\big\rangle
    &=\big\langle(J^a_1\mathcal O_i)(\infty)\mathcal O_j(1)\mathcal O_k(0)\big\rangle
    -\big\langle\mathcal O_i(\infty)(J^a_0\mathcal O_j)(1)\mathcal O_k(0)\big\rangle\notag\\[1ex]
    &\quad+\big\langle\mathcal O_i(\infty)(J^a_1\mathcal O_j)(1)\mathcal O_k(0)\big\rangle .
    \label{eq:Ward-identities_1}
\end{align}
Next, we consider
\begin{align}
    \big\langle\mathcal O_i(\infty)(J^a_{-1}\mathcal O_j)(1)\mathcal O_k(0)\big\rangle=\oint_1\frac{dz}{2\pi i}\frac{1}{z-1}\big\langle\mathcal O_i(\infty)J^a(z)\mathcal O_j(1)\mathcal O_k(0)\big\rangle.
\end{align}
Deforming this contour to $z=0$ and $z=\infty$, we first expand the kernel near $z=0$:
\begin{align}
    \frac{1}{z-1}=-\frac{1}{1-z}=-1-z-z^2-\cdots .
\end{align}
Hence
\begin{align}
    \oint_{0}\frac{dz}{2\pi i}\,
    \frac{1}{z-1}J^a(z)=-J^a_0-J^a_1-\cdots .
\end{align}
The contour deformation gives another minus sign, and therefore this part contributes
\begin{align}
    J^a_0+J^a_1 .
\end{align}
For the contour at infinity,
\begin{align}
    \oint_\infty\frac{dz}{2\pi i}\frac{1}{z-1}J^a(z)
    &=-\oint_{w=0}\frac{dw}{2\pi i}\frac{w}{1-w}J^a(w)\notag\\[1ex]
    &=-J^a_1-J^a_2-\cdots,
\end{align}
Again using \eqref{eq:state-condition}, only the $J^a_1$ term survives after the contour deformation. Thus
\begin{align}
    \big\langle\mathcal O_i(\infty)(J^a_{-1}\mathcal O_j)(1)\mathcal O_k(0)\big\rangle
    &=\big\langle(J^a_1\mathcal O_i)(\infty)\mathcal O_j(1)\mathcal O_k(0)\big\rangle
    +\big\langle\mathcal O_i(\infty)\mathcal O_j(1)(J^a_0\mathcal O_k)(0)\big\rangle\notag\\[1ex]
    &\quad+\big\langle\mathcal O_i(\infty)\mathcal O_j(1)(J^a_1\mathcal O_k)(0)\big\rangle .
    \label{eq:Ward-identities_2}
\end{align}
Finally, we consider the function at infinity. In the local coordinate $w=1/z$, the mode $J^a_{-1}$ at infinity is represented by the kernel $w^{-1}=z$. Therefore the relevant contour can be written as a contour at infinity with kernel $zJ^a(z)$. Deforming it to the finite insertions gives
\begin{align}
    \oint_{\infty}\frac{dz}{2\pi i}zJ^a(z)=\oint_{0}\frac{dz}{2\pi i}zJ^a(z)=J^a_1
\end{align}
and, near $z=1$,
\begin{align}
    \oint_{\infty}\frac{dz}{2\pi i}zJ^a(z)=\oint_{1}\frac{dz}{2\pi i}(1+z-1)J^a(z)=J^a_0+J^a_1,
\end{align}
Consequently,
\begin{align}
    \big\langle(J^a_{-1}\mathcal O_i)(\infty)\mathcal O_j(1)\mathcal O_k(0)\big\rangle
    &=\big\langle\mathcal O_i(\infty)\mathcal O_j(1)(J^a_1\mathcal O_k)(0)\big\rangle
    +\big\langle\mathcal O_i(\infty)(J^a_0\mathcal O_j)(1)\mathcal O_k(0)\big\rangle\notag\\[1ex]
    &\quad+\big\langle\mathcal O_i(\infty)(J^a_1\mathcal O_j)(1)\mathcal O_k(0)\big\rangle .
    \label{eq:Ward-identities_3}
\end{align}
We first compute $C_{\phi^{UV}\phi^{UV}\phi^{UV}_w}$. Since
\begin{align}
    J^0_0\Phi^{UV}=J^0_0\Phi_0=0,
\end{align}
the $J^0_{-1}\Phi_0$ term in \eqref{eq:D-state} does not contribute. Using the Ward identities \eqref{eq:Ward-identities_1}, we find
\begin{align}
    \big\langle\Phi^{UV}(\infty)\Phi^{UV}(1)(J^+_{-1}\Phi_{-1})(0)\big\rangle
    &=-\big\langle\Phi_0(\infty)(J^+_0\Phi_0)(1)\Phi_{-1}(0)\big\rangle\notag\\[1ex]
    &=-\sqrt 2\,K_N
    \begin{pmatrix}
        1 & 1 & 1\\
        0 & 1 & -1
    \end{pmatrix}\notag\\[1ex]
    &=-\frac{1}{\sqrt 3}K_N .
    \label{eq:three-point-1}
\end{align}
Similarly,
\begin{align}
    \big\langle\Phi^{UV}(\infty)\Phi^{UV}(1)(J^-_{-1}\Phi_{1})(0)\big\rangle=\frac{1}{\sqrt 3}K_N .
    \label{eq:three-point-2}
\end{align}
Therefore,
\begin{align}
    C_{\phi^{UV}\phi^{UV}\phi^{UV}_w}=\sqrt{\frac{N}{2(N-2)(N+4)}}\frac{2}{\sqrt 3}K_N .
    \label{eq:CLLw}
\end{align}
This is a chiral structure constant in the phase convention fixed by the state in \eqref{eq:D-state}. The factorwise UV chiral charge-conjugation automorphism used to characterize the $W^{(3)}_0$ charge is not assumed to be an unbroken global symmetry of the folded RG-wall OPE, and therefore it does not impose a vanishing selection rule on this chiral coefficient.
The coefficient $C_{\phi^{UV}_w\phi^{UV}_w\phi^{UV}_w}$ can be computed in the same manner. Using the definition of $\phi^{UV}_w$ in terms of $v_\pm,\,v_0$, one finds
\begin{align}
    C_{\phi^{UV}_w\phi^{UV}_w\phi^{UV}_w}=\left(\frac{N}{2(N-2)(N+4)}\right)^{\frac{3}{2}}(-C_{v_+v_+v_+}+C_{v_+v_+v_-}+\cdots)
\end{align}
where the ellipsis denotes the remaining terms obtained by expanding the three factors of $\phi_{w}^{UV}$. By repeatedly applying the affine algebra \eqref{eq:A1-alg} together with the Ward identities \eqref{eq:Ward-identities_1}, \eqref{eq:Ward-identities_2}, and \eqref{eq:Ward-identities_3}, each term can be reduced to a finite linear combination of the basic three-point functions of the form \eqref{eq:three-point-1} and \eqref{eq:three-point-2}. 

As an illustration, let us consider the term $C_{v_+v_+v_-}$. Throughout the following computation we fix the insertion points to be $z_1=\infty$, $z_2=1$, and $z_3=0$, and suppress them from the notation. We obtain
\begin{align}
    C_{v_+v_+v_-}&=\big\langle J^+_{-1}\Phi_{-1},v_+,v_-\big\rangle\notag\\[1ex]
    &=\big\langle\Phi_{-1},J^+_0v_+,v_-\big\rangle+\big\langle\Phi_{-1},J^+_1v_+,v_-\big\rangle+\big\langle\Phi_{-1},v_+,J^+_1v_-\big\rangle\notag\\[1ex]
    &=\sqrt{2}\big\langle\Phi_{-1},J^+_{-1}\Phi_0,J^-_{-1}\Phi_1\big\rangle+(N+2)\big\langle\Phi_{-1},J^+_{-1}\Phi_{-1},\Phi_1\big\rangle\notag\\[1ex]
    &=\sqrt{2}\bigl( -\big\langle\Phi_{-1},J^-_0J^+_{-1}\Phi_0,\Phi_1\big\rangle+\big\langle\Phi_{-1},J^-_1J^+_{-1}\Phi_0,\Phi_1\big\rangle \bigr)+(N+2)\big\langle\Phi_{-1},\Phi_{-1},J^+_0\Phi_1\big\rangle\notag\\[1ex]
    &=\sqrt{2}\bigl( 2\big\langle\Phi_{-1},J^0_{-1}\Phi_0,\Phi_1\big\rangle-\sqrt{2}\big\langle\Phi_{-1},J^+_{-1}\Phi_{-1},\Phi_1\big\rangle+N\big\langle\Phi_{-1},\Phi_0,\Phi_1\big\rangle \bigr)\notag\\[1ex]
    &=\sqrt{2}(N+2)\big\langle\Phi_{-1},\Phi_0,\Phi_1\big\rangle\notag\\[1ex]
    &=\frac{N+2}{\sqrt{3}}K_N
\end{align}
Carrying out the same reduction for all the remaining terms in the expansion, we finally obtain
\begin{align}
    C_{\phi^{UV}_w\phi^{UV}_w\phi^{UV}_w}=\sqrt{\frac{N}{2(N-2)(N+4)}}\frac{\sqrt 6\,(N-4)}{2(N-2)}K_N .
    \label{eq:Cwww}
\end{align}
It remains to include the minimal-model factor. The field $\phi^{IR}$ also corresponds to level-one $\mathfrak{su}(2)_N$ sector. With the same normalization convention for $K_N$, its three-point coefficient is
\begin{align}
    C_{\phi^{IR}\phi^{IR}\phi^{IR}}=\frac{2\sqrt 3\,(N-2)}{\sqrt{N(N-1)}\,K_N}.
    \label{eq:CRRR}
\end{align}
Combining \eqref{eq:Cwww} and \eqref{eq:CRRR}, and using $X=\phi_{w}^{UV}\phi^{IR}$, we obtain
\begin{align}
    C_{XXX}
    =C_{\phi^{UV}_w\phi^{UV}_w\phi^{UV}_w}C_{\phi^{IR}\phi^{IR}\phi^{IR}}=\frac{3\sqrt2(N-4)}{\sqrt{(N-1)(N-2)(N+4)}}.
    \label{eq:CXXX}
\end{align}
The coefficient $C_{W^-W^-X}$ factorizes in the same way into the left and right sectors. From the definition of $W^-$, the relevant relative coefficient is
\[
    \frac{h^{UV}h^{IR}}{2h^{UV}h^{IR}}=\frac{1}{2}.
\]
Therefore,
\begin{align}
    C_{W^-W^-X}
    =\frac{1}{2}C_{\phi^{UV}\phi^{UV}\phi^{UV}_w}C_{\phi^{IR}\phi^{IR}\phi^{IR}}\notag=\sqrt{\frac{2(N-2)}{(N-1)(N+4)}} .
    \label{eq:CWWX}
\end{align}
Finally, the OPE between $X$ and $W^-$ generates $J$. To determine the corresponding coefficient, we consider $C_{XW^-J}$. The relevant factor is 
\[
    i\sqrt{\frac{h^{IR}}{2h^{UV}}}=i\sqrt{\left(\frac{N}{2+N}\right)\left(\frac{2+N}{4}\right)}=i\sqrt{\frac{N}{4}} ,
\]
and hence
\begin{align}
    C_{XW^-J}
    =i\sqrt{\frac{N}{4}}C_{\phi^{UV}_w\phi^{UV}\phi^{UV}}C_{\phi^{IR}\phi^{IR}\phi^{IR}}=i\sqrt{\frac{2N(N-2)}{(N-1)(N+4)}} .
    \label{eq:CXWJ}
\end{align}

\bibliographystyle{JHEP}

\newpage
\bibliography{refs.bib}

\end{document}